\documentclass[aps,prl,twocolumn,superscriptaddress]{revtex4-1}

\usepackage{graphicx}
\usepackage{amssymb}
\usepackage{amsmath}
\usepackage{color}
\usepackage{physics}
\usepackage{bbm}
\usepackage{hyperref}
\usepackage{epstopdf}
\usepackage{pdfpages}

\newcommand{\h}{\hat}
\newcommand{\Size}{\mathcal{S}}
\newcommand{\err}{\varepsilon}
\newcommand{\Norm}{\mathcal{N}}
\renewcommand{\d}{\partial}
\newcommand{\AvgSize}{\overline{\Size}}

\newcommand{\SizeWidth}{\delta \Size}

\makeatletter
\AtBeginDocument{\let\LS@rot\@undefined}
\makeatother

\begin{document}

\title{Operator Growth  in Open Quantum Systems}

\author{Thomas Schuster}
\affiliation{Department of Physics, University of California, Berkeley, California 94720 USA}
\author{Norman Y. Yao}
\affiliation{Department of Physics, University of California, Berkeley, California 94720 USA}
\affiliation{Materials Science Division, Lawrence Berkeley National Laboratory, Berkeley, California 94720, USA}
\date{\today}


\newcommand{\TableRegimes}{
\begin{table}[t]
\vspace{-2.3mm}
\caption{Size distributions in various physical regimes} 
\centering 
\begin{tabular}{c c c} 
\hline\hline 
System & \,\,\,\, Unitary dynamics \,\,\,\, &  Open dynamics
\\ [0.5ex]
\hline 
$(d\geq1)$D, no & peaked-size  & size unaffected\\[-0.5ex]
conservation law& $\AvgSize \sim t^d$ & $\AvgSize \sim t^d$ \\[-1ex]
&  & \\[-1.7ex]
$(d\geq1)$D, con-  &  bimodal & size decreases \\[-0.5ex]
served quantity & $\AvgSize \sim t^d$ & $\AvgSize \sim 1/t^d$ \\[-1ex]
&  & \\[-1.7ex]
all-to-all &  broad size & size plateaus \\[-0.5ex]
coupled (0D) & $\AvgSize \sim e^{\lambda t}$ & $\AvgSize \sim \lambda/\err$ \\[-1ex]
&  & \\[-1.7ex]
$(d\geq1)1$D,  &  broad size  & size growth slowed \\[-0.5ex]
long-range & super-ballistic & $\AvgSize \sim t^d$ \\[-1ex]
&  & \\[-1.7ex]
free fermion &  broad size & size decreases \\[-0.5ex]
integrable & $\AvgSize \sim t$ & $\AvgSize \sim 1/t$ \\[+1ex]
\hline 
\vspace{-3.3mm}
\end{tabular}
\label{table}
\end{table}
}

\newcommand{\FigureOne}{
\begin{figure}
\centering
\includegraphics[width=\columnwidth]{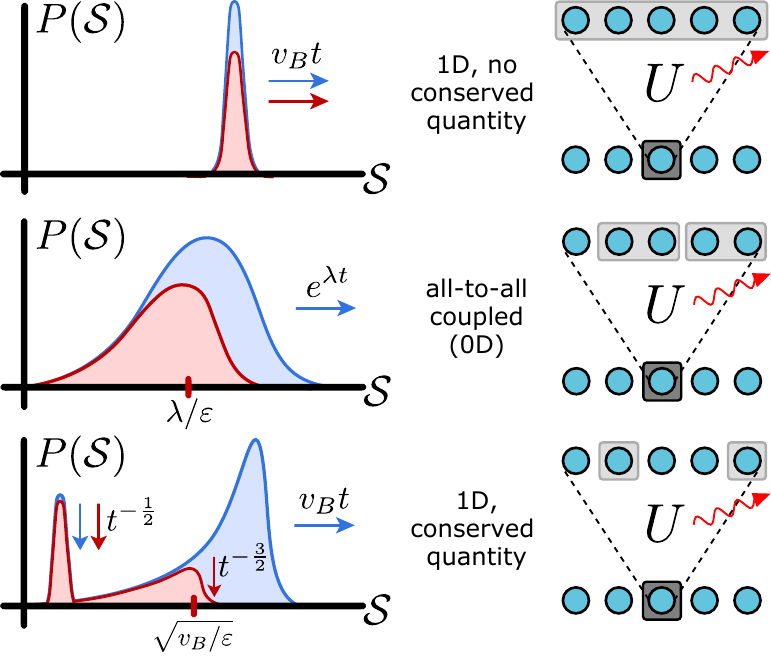}
\caption{
Left: Size distributions for three classes of systems under unitary (blue) versus open (red) dynamics.
Rightward arrows denote growth in time to larger sizes, ticks denote a fixed size, and downward arrows denote loss of probability at a given size.
Right: Qualitative depiction of open-system operator growth.
In all cases, operators lose normalization due to open dynamics (dark to light gray boxes).
In the latter two classes, operators are dominated by smaller size components compared to unitary evolution (smaller boxes).
} 
\label{fig: 1}
\end{figure}
}

\newcommand{\FigureOneZero}{
\begin{figure}
\centering
\includegraphics[width=\columnwidth]{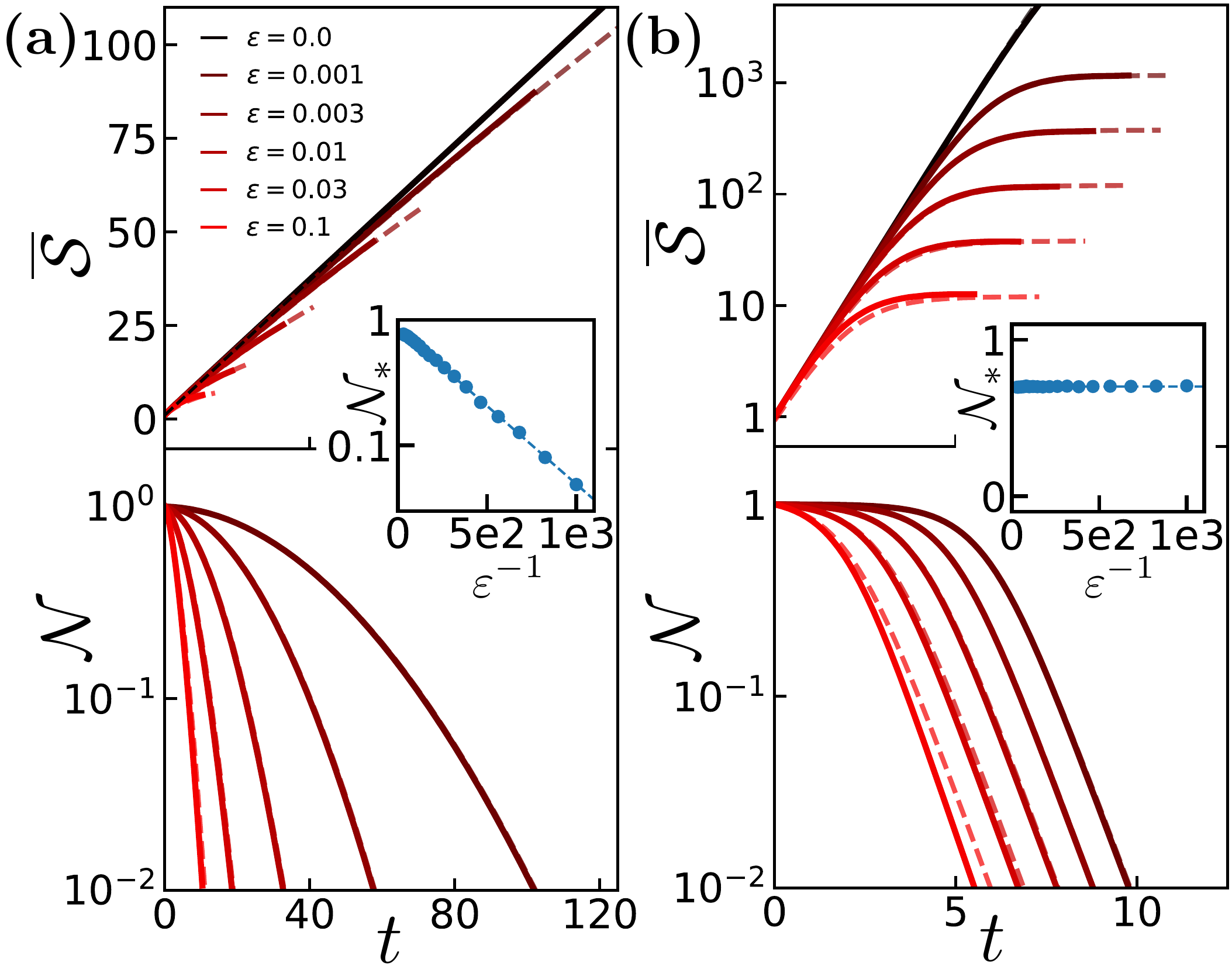}
\caption{\textbf{(a)} Average operator size, $\overline{\Size}$, and the Loschmidt echo fidelity, $\mathcal{N}$, in a 1D RUC with $N =  200$~\cite{suppinfo}. 
The size grows ballistically with quadratic corrections due to open-system dynamics (solid, data; dashed, theory). 
Inset: The Loschmidt echo fidelity, $\mathcal{N_*}$, when $\frac{d\overline{\Size}}{dt} = 0.9 \frac{d\overline{\Size}}{dt}\big|_{\err=0}$, decays exponentially in the inverse error rate, $\err^{-1}$. %
\textbf{(b)} All-to-all RUC with $N = 1500$~\cite{suppinfo}. The size grows exponentially before plateauing to a value which is independent of the system size. The decay rate of the Loschmidt echo is independent of  $\err$ after plateauing (solid, data; dashed, theory).
Inset: The Loschmidt echo fidelity, $\mathcal{N_*}$, when $\frac{d\log\overline{\Size}}{dt} = 0.9 \frac{d\log\overline{\Size}}{dt}\big|_{\err=0}$, is constant with respect to $\err$.
} 
\label{fig: numerics 1}
\end{figure}
}

\newcommand{\FigureNumerics}{
\begin{figure}
\centering
\includegraphics[width=\columnwidth]{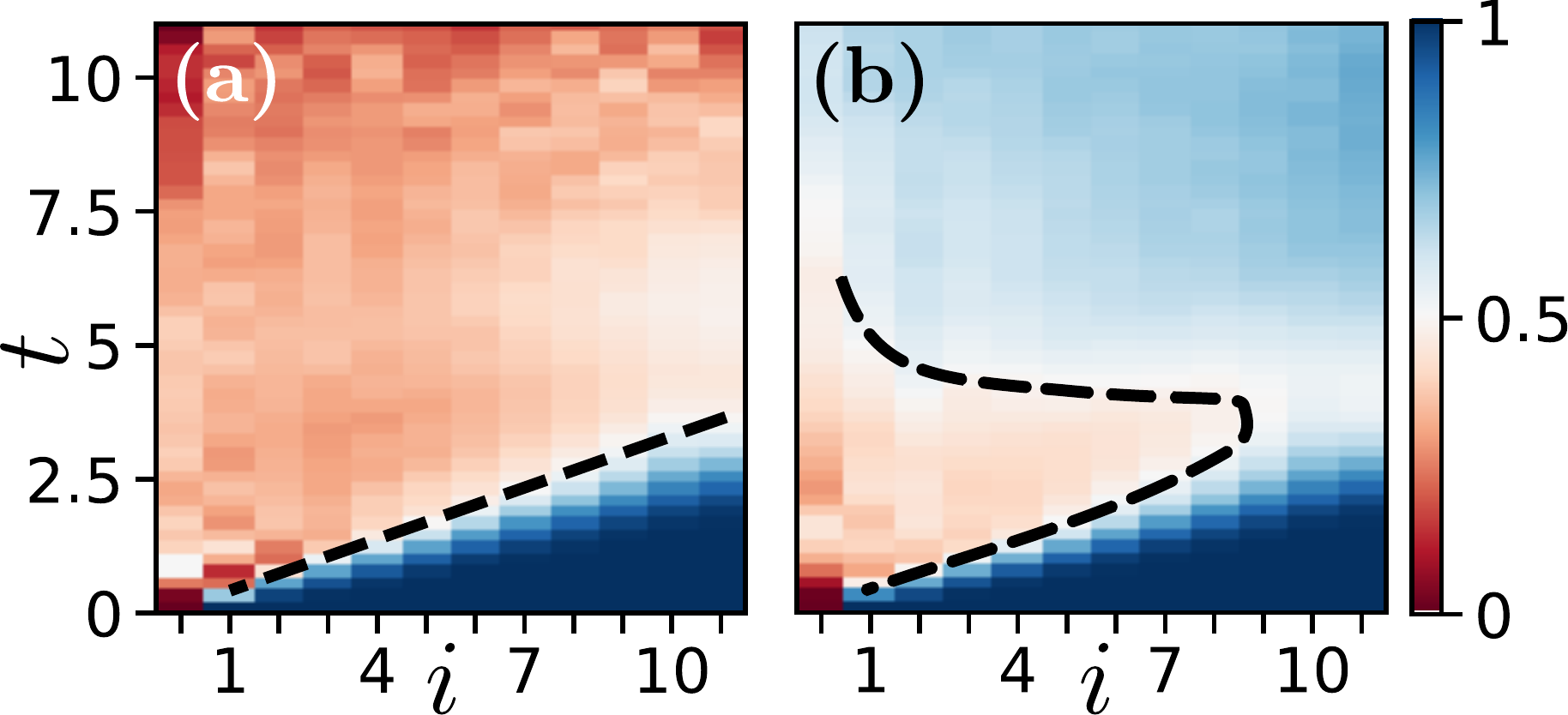}
\caption{The OTOC as a function of time and space for an $N=12$ one-dimensional spin chain. \textbf{(a)} Operators that do not overlap the Hamiltonian exhibit an OTOC which follows a ballistic light cone. \textbf{(b)} For an operator that overlaps with the Hamiltonian, the OTOC at a given site $i$ initially decays, before increasing at later times. 
The specific OTOC we calculate takes the form, $\frac{1}{4} \sum_{P} \langle e^{-iH_1t} \h{M} e^{iH_1t}  \h{P}_i e^{-iH_2t} \h{M} e^{iH_2t} \h{P}_i \rangle/ \mathcal{N}(t)$, where the forwards and backwards time-evolution are governed by two distinct 1D Hamiltonians, $H_1 = H_2 + \eta \, \delta H$ (for details see supplemental materials~\cite{suppinfo}).
} 
\label{fig: numerics 2}
\end{figure}
}

\newcommand{\FigureCircuit}{
\begin{figure}
\centering
\includegraphics[width=\columnwidth]{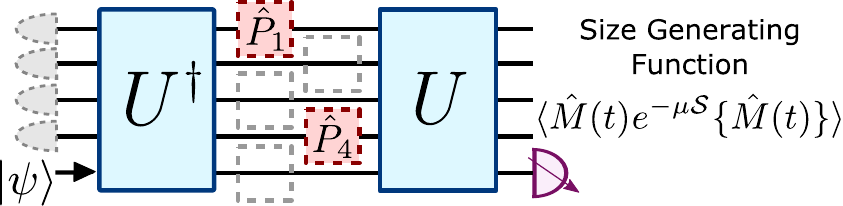}
\caption{
Protocol to measure the  generating function, $G_\Size(\mu)$, of the operator size distribution. Gray qubits are initially random in the computational basis, $\ket{\psi}$ is an $\hat{M}$ eigenstate, and each site is acted upon randomly by either the identity (gray) or a non-identity Pauli operator (red) in each experimental shot. 
} 
\label{fig: circuit}
\end{figure}
}

\begin{abstract}
The spreading of quantum information in closed systems, often termed scrambling, is a hallmark of many-body quantum dynamics.
In open systems, scrambling competes with noise, errors and decoherence.
Here, we provide a universal framework that describes the scrambling of quantum information in open systems:  we predict that the effect of open-system dynamics is fundamentally controlled by operator size distributions and independent of the microscopic error mechanism.  
This framework allows us to demonstrate that open quantum systems exhibit universal classes of information dynamics that fundamentally differ from their unitary counterparts.
Implications for Loschmidt echo experiments and the classical simulability of open quantum dynamics will be discussed. 
\end{abstract}

\maketitle

Conventionally, the study of quantum many-body systems has focused on the prediction of few-body observables, such as local correlation functions. 
More recently, sparked by fundamental questions in quantum thermalization and chaos~\cite{bohrdt2017scrambling}, the classical simulation of quantum systems~\cite{verstraete2004matrix}, and quantum gravity~\cite{harlow2016jerusalem}, physicists have turned to a complementary pursuit: quantifying the complexity of many-body   \emph{dynamics} itself.

At the heart of this pursuit is the notion of quantum information \emph{scrambling}; across nearly the entirety of interacting many-body  quantum systems, information encoded in initially local operators grows to become highly non-local~\cite{lieb1972finite,roberts2015localized}.
Remarkably, recent experimental advances have enabled the direct measurement of scrambling---a task that, most commonly, utilizes backwards time-evolution~\cite{li2017measuring,garttner2017measuring,sanchez2019emergent,sanchez2021emergent,dominguez2021decoherence,dominguez2021dynamics,mi2021information,cotler2022information}, but can also be performed using multiple copies of the system~\cite{islam2015measuring,landsman2019verified,blok2020quantum} or randomized measurements~\cite{brydges2019probing,joshi2020quantum}.
In such systems, the interplay between scrambling dynamics, extrinsic decoherence, and  experimental noise motivates an essential  question: What is the nature of quantum information scrambling in \emph{open} quantum systems~\cite{landsman2019verified,arute2019quantum,alonso2019out,mi2021information,yoshida2019disentangling,swingle2018resilience,vermersch2019probing,zhang2019information,agarwal2020toy,bao2021hayden,touil2021information,andreadakis2022scrambling,harris2022benchmarking}?

In this Letter, we introduce a universal framework---based upon \emph{operator size distributions}~\cite{nahum2018operator,roberts2018operator,qi2019quantum,schuster2021many}---for capturing the effect of local errors on scrambling dynamics.
In particular, we conjecture that the propagation of errors in chaotic many-body systems is fundamentally controlled by the size distributions of time-evolved  operators and independent of the microscopic error mechanism. 
Our framework immediately offers predictions for both the Loschmidt echo~\cite{loschmidt1876volume,goussev2012loschmidt} and out-of-time-ordered correlation (OTOC) functions~\cite{larkin1969quasiclassical,xu2022scrambling}.
In particular, we predict that  the decay of the Loschmidt echo, which measures the fidelity associated with backwards time-evolution, occurs at a rate proportional to the operator size. 
Meanwhile, we predict that the decay of the OTOC, which measures the growth of  local operators, is inhibited by open-system dynamics (by an amount proportional to the width of the operator size distribution).

\FigureOne

We leverage our framework to characterize operator growth in five distinct classes of open quantum systems, which vary in their dimensionality, range of interaction, conservation laws, and integrability (Table~\ref{table}, Figs.~\ref{fig: 1}--\ref{fig: numerics 2}).
In each class, our framework yields markedly distinct predictions for  the Loschmidt echo  and OTOCs.
We hypothesize that these results provide a theoretical underpinning for  recent nuclear magnetic resonance (NMR) experiments~\cite{sanchez2019emergent,sanchez2021emergent,dominguez2021decoherence}, and also serve to resolve apparent disagreements between previous empirical studies of open-system scrambling~\cite{vermersch2019probing,mi2021information,zhang2019information}.
Finally, we propose and analyze a protocol for measuring operator size distributions via engineered dissipation.

\TableRegimes

\emph{Operator size distributions.}---We begin with a simple example to build intuition.
Consider a lattice of qubits acted on by a series of local quantum gates, each featuring some error $\varepsilon$, before measuring a local operator $\hat{M}$.
Noting that the measurement can only be influenced by gates in its past light cone, a naive estimate of the measurement fidelity is $\mathcal{F} \approx (1-\varepsilon)^{\mathcal{V}_{\text{LC}}}$, where $\mathcal{V}_{\text{LC}}$ is the light cone volume, i.e.~the number of gates it contains~\cite{arute2019quantum,mi2021information}.
This relation in fact already contains the essential intuition underlying our work: a connection between the measurement fidelity of a local operator and the operator's growth under Heisenberg evolution.
By generalizing the light cone volume using operator size distributions, we will show that this connection is significantly richer and more universal than the above example suggests~\cite{roberts2015localized,roberts2018operator}.

To introduce the notion of an operator size distribution, we first define the size of a Pauli string, $\h{R}$, as its number of non-identity elements; for instance, $\hat{R} = Y \otimes \mathbbm{1} \otimes Z \otimes X$ has size $\Size_R = 3$.
From this, one can define the size superoperator:
\begin{equation} \label{superoperator}
\Size\{ \h{\mathcal{O}} \} \equiv - \sum_{P_i} ( P_i \h{\mathcal{O}} P_i^\dagger - \h{\mathcal{O}})/4,
\end{equation}
which gives $\Size\{ \h{R} \} = \Size_{R} \h{R}$, where $\h{P}_i \in \{ \h{\mathbbm{1}}_i, \h{X}_i, \h{Y}_i, \h{Z}_i \}$ are single-qubit Pauli operators~\cite{schuster2021many}.
More general operators can be expressed as a sum of Pauli strings, $\h{\mathcal{O}} = \sum_{\h{R}} c_{R} \h{R}$, and thereby possess a \emph{size distribution}, $P(\Size) = \sum_{\{ \Size_{R} = \Size\} } |c_{R}|^2$, with normalization  $\Norm = \langle \mathcal{O}^\dagger \mathcal{O} \rangle = \sum_R |c_R|^2$; here, $\langle \cdot \rangle \equiv \Tr( \cdot )/\Tr(\mathbbm{1})$  represents the infinite temperature expectation value.
We note that the operator size distribution is closely related to out-of-time-ordered correlation functions, $\langle \h{M}(t) \h{V}_j \h{M}(t) \h{V}_j \rangle$.
As an operator, $\h{M}(t)$, grows to have support on site $j$, the OTOC typically decays to zero.
From Eq.~(\ref{superoperator}), one immediately sees that the average size of $\h{M}(t)$ is directly proportional to unity minus the OTOC averaged over all single-qubit Pauli operators:
\begin{equation} \label{avgsize}
 \overline{\Size} = \frac{\langle \h{M}(t) \Size \{ \h{M}(t) \} \rangle}{\langle \h{M}(t)  \h{M}(t) \rangle} =  \frac{1}{4} \sum_{P_i} \left ( 1 -  \frac{\langle \hat{M}(t) P_i \hat{M}(t) P_i  \rangle}{\langle \hat{M}(t) \hat{M}(t)  \rangle} \right ).
 \end{equation}

\emph{Open-system operator growth hypothesis.}---Let us now turn to open quantum systems. Operator evolution is typically governed by the Lindblad master equation:
\begin{equation} \label{Lindbladian}
\d_t \h{M} =  i [\h{H}, \h{M}] -  \sum_\alpha \err_\alpha \left( \h{L}_\alpha^\dagger \h{M} \h{L}_\alpha - \frac{1}{2} \{ \h{L}^\dagger_\alpha \h{L}_\alpha , \h{M} \} \right).
\end{equation}
The first term describes unitary time-evolution, while the second describes a sum of local error processes, each characterized by a Lindblad operator, $\h{L}_\alpha$, and an associated error rate, $\err_\alpha$.

Our central conjecture is that the effects of local errors on operator growth are in fact captured by a much simpler, effective Lindblad equation:
\begin{equation} \label{Lindbladian size}
\d_t \h{M} =  i [\h{H}, \h{M}] - \err \Size \{ \h{M} \},
\end{equation}
where $\Size$ is the size superoperator.
In effect, this model replaces the original Lindblad operators with isotropic decoherence at each qubit [Eq.~(\ref{superoperator})].

This conjecture is rooted in the following intuition---higher size operators are affected by a greater number of local error processes, and thus decohere at a faster rate~\cite{fn3}.
More specifically, we expect large-size components of $\h{M}$ to typically involve exponentially many Pauli strings varying rapidly in time.
This  serves to ``average'' the effect of Lindblad operators such that their action depends solely on whether they are in the support  of a given size component of $\h{M}$, independent of their precise microscopic form.
The number of Lindblad operators in the support is directly proportional to the size.

Our framework predicts two effects of open-system dynamics on operator growth, which are captured by the behavior of the Loschmidt echo and  the average OTOC (i.e.~the average operator size), respectively. 
For the former,  we note that the Loschmidt echo fidelity with respect to a local operator  is in fact equal to the \emph{normalization} of the operator's size distribution, $\Norm (t) = \langle \h{M}(t) \h{M}(t) \rangle = \int d\Size P(\Size)$. 
Our framework predicts~\cite{fn5} that the Loschmidt echo decays in time at a rate equal to the average size multiplied by the error rate:
\begin{equation} \label{dt norm}
\partial_t \log \Norm(t) = - 2 \err \AvgSize (t).
\end{equation}

Turning to the OTOC,  we note that errors decrease the amplitude of large-size components of $\h{M}(t)$ at a faster rate than small-size components.
Thus, compared to purely unitary evolution, open-system dynamics inhibit the growth of operators.
More specifically, we predict that the average size, $\AvgSize$ [related to the OTOC via Eq.~(\ref{avgsize})], evolves according to:
\begin{equation}\label{dt size}
\partial_t \AvgSize (t) = (\text{unitary}) - 2 \err \SizeWidth(t)^2.
\end{equation}
Here, the first term captures the specific unitary dynamics of the system, while the second term decreases the size at a rate proportional to  the variance of the size distribution, $\delta \Size^2$.

\emph{Open-system scrambling dynamics.}---We now apply our framework to five distinct classes of scrambling dynamics (Table~\ref{table}, Figs.~\ref{fig: 1}--\ref{fig: numerics 2}): local and all-to-all interacting systems without conservation laws, local systems \emph{with} conservation laws, long-range interacting systems, and free fermion integrable systems (see supplemental materials for a detailed discussion of the latter two cases~\cite{suppinfo}).

We begin by demonstrating that operator growth in two paradigmatic scramblers---systems with no conserved quantities under local and all-to-all interactions---are affected by open-system dynamics in drastically different ways. 
For the former (focusing on 1D systems for specificity), one expects operators to grow ballistically in time under unitary dynamics, with $\AvgSize \approx \frac{3}{2} v_B t$, where $v_B$ is the butterfly velocity.
Meanwhile, the width of the operator size distribution grows ``diffusively'', $\delta \Size \approx c \sqrt{v_B t}$ where $c$ is a constant~\cite{nahum2018operator,von2018operator,schuster2021many}.

Combining these expectations via Eq.~(\ref{dt size}), we arrive at a simple phenomenological equation for operator growth under \emph{open-system} dynamics, $\d_t \AvgSize \approx \frac{3}{2} v_B - \err (c \sqrt{v_B t})^2$, whose solution yields the prediction: $\AvgSize (t) \approx \frac{3}{2} v_B t - \frac{c}{2} \err  v_B t^2$.
From Eq.~(\ref{dt norm}), the Loschmidt echo fidelity thus decays as a Gaussian in time, $\Norm(t) = \exp \big( - \err \int_0^t dt' \, \overline{\Size}(t') \big) \approx \exp( - \frac{3}{4} \err v_B t^2 )$, to leading order in $\varepsilon$.
To explore these predictions, we numerically simulate the open-system dynamics of a 1D random unitary circuit (RUC)~\cite{nahum2017quantum,nahum2018operator,suppinfo}.
As depicted in Fig.~\ref{fig: numerics 1}(a), we find that both the operator size and the Loschmidt echo fidelity (solid lines) agree remarkably well with our phenomenological predictions (dashed lines) across multiple orders of magnitude in the error rate.

In contrast, in all-to-all interacting systems,  unitary dynamics typically exhibit ``fast scrambling'' characterized by the exponential growth of operator size in time, $\AvgSize \sim \! e^{\lambda t}$, where $\lambda$ is the Lyapunov exponent~\cite{sachdev1993gapless,kitaev2015simple,maldacena2016remarks,kobrin2020many,schuster2021many,bentsen2019fast}.
Unlike local systems, the size distribution is also extremely broad, $\SizeWidth \approx b \Size$ where $b$ is a constant, owing to the exponential growth of early-time fluctuations~\cite{roberts2018operator,qi2019quantum,schuster2021many}.
Solving Eq.~(\ref{dt size}), i.e. $\partial_t \AvgSize \approx \lambda \AvgSize - \err b^2 \AvgSize\hspace{0.1mm}^2$, then yields an intriguing prediction: under open-system dynamics, the average operator size \emph{plateaus} to a system-size independent value, $\AvgSize_p \approx \lambda / (\err b^2)$, after a time $t_p \sim \log( \lambda / (\err b^2) )$.
This  causes the Loschmidt echo to approach a constant rate of decay, $\Norm(t) \sim \exp( - \lambda t / b^2 )$.
Notably, the decay rate, $\lambda  / b^2$, is \emph{independent} of the microscopic error rate, $\err$, echoing seminal results in single-particle quantum chaos~\cite{jalabert2001environment} and tantalizing recent NMR experiments~\cite{sanchez2019emergent,sanchez2021emergent,dominguez2021decoherence}.
As shown in Fig.~\ref{fig: numerics 1}(b), both of these predictions are indeed born out by RUC simulations.

\FigureOneZero

One can further sharpen the distinction between open-system dynamics for local versus all-to-all interactions, by analyzing their behavior at asymptotically small error rates.
Specifically, consider the value of the Loschmidt echo, $\mathcal{N_*}$, at a time when the open-system dynamics have substantially deviated from the unitary dynamics.
In all-to-all systems, this occurs shortly after the plateau time, $t_p$, which gives an order one Loschmidt echo, $\Norm(t_p) \approx \exp \big( -\err \int_0^{t_p} dt' \, e^{\lambda t'} \big) \approx \exp( - 1 / b^2 )$, independent of the error rate [inset, Fig.~\ref{fig: numerics 1}(b)].

In contrast, in 1D systems this  occurs  when $(\err v_B t^2 / v_B t) \sim 1$, at which point the Loschmidt echo has decayed to an \emph{exponentially} small value, $\Norm(t) \sim \exp( - v_B / \err)$ [inset, Fig.~\ref{fig: numerics 1}(a)].
For small error rates, this implies that large deviations in operator growth are in practice unobservable for locally-interacting systems, since the  signal is exponentially small in $1/\err$.
Physically, this is a direct consequence of the asymptotic separation, $\delta \Size \ll \Size$.

\emph{Effects of conservation laws.}---We now show that the above behaviors are strikingly modified when an operator has overlap with a conserved quantity, $\h{Q} = \sum_i \h{q}_i$ (e.g.~the total spin, or the Hamiltonian).
Such systems feature an interplay between hydrodynamics and scrambling, which is embodied by a `bimodal' profile for unitary time-evolved operators~\cite{khemani2018operator,rakovszky2018diffusive}:
\begin{equation} \label{conserved operator}
\h{M}(t) = \sum_i q(i,t) \, \h{q}_i + \sum_{\h{R} \neq \h{q}_i}  c_R(t) \h{R}.
\end{equation}
The operator contains both small-size components, $\h{q}_i$, representing the dynamics of the conserved quantity,  as well as  large-size Pauli strings, $\h{R}$, representing scrambled information.

The small-size components arise because an operator's overlap with $\h{Q}$, $\langle\h{M}(t) \h{Q} \rangle = \int dx \, q(x,t)$, is conserved in time.
As an example, in chaotic 1D systems,  one expects the local overlap,  $q(i,t) = \langle \h{M} (t) \h{q}_i \rangle$, to spread diffusively, which causes the total normalization of the small-size components to decay in time, $\int dx \, |q(x,t)|^2 \sim \! 1/\sqrt{t}$.
This in turn, implies that the total normalization of the large-size Pauli strings is increasing in time; physically, this corresponds to the dynamics of $q(i,t)$ `emitting' chaotic components, which  spread ballistically from thereon.
In combination, this leads to a size distribution [Fig.~\ref{fig: 1}], $P(\Size) \approx \frac{1}{\sqrt{D t}} \delta(\Size-1) + \frac{v_B}{\sqrt{D}}(\frac{3}{2} v_B t - \Size)^{-3/2}$, where we have assumed that $\Size_{q_i} = 1$~\cite{rakovszky2020dissipation}.

\FigureNumerics

We expect open-system dynamics to damp the large-size components of $\h{M}$ by a factor $\sim \! e^{-\err \Size^2 / v_B}$, where $\Size^2 / v_B$ characterizes the space-time volume of a chaotic component~\cite{fn4}. 
This effectively truncates the size distribution above $\Size_{\text{tr}} \! \sim \! \sqrt{v_B/\err}$ [Fig.~\ref{fig: 1}].
At late times, $v_B t \gtrsim \Size_{\text{tr}}$, this suggests that the average operator size will actually \emph{shrink} in time, since small-size components decay more slowly $\sim \! t^{-1/2}$, than large-size components, $P( \Size_{\text{tr}} ) \sim \! t^{-3/2}$.
This sharply contrasts with the behavior of operators that do not overlap conserved quantities, where one expects monotonic growth [Fig.~\ref{fig: numerics 1}(a)].

To explore this, we simulate the dynamics of a one-dimensional spin chain and measure the OTOC as a proxy for operator growth. 
For an operator that does not overlap with the Hamiltonian, we find that the OTOC  decays monotonically following a linear light-cone [Fig.~\ref{fig: numerics 2}(a)]. 
For an operator exhibiting overlap, we find that the decay of the OTOC indeed \emph{reverses} as a function of time, indicative of a decrease in the average operator size [Fig.~\ref{fig: numerics 2}(b)].
Interestingly, this insight immediately resolves an apparent disagreement between previous studies of open-system operator growth. In particular, certain studies found that OTOCs were only minimally affected by errors~\cite{vermersch2019probing,mi2021information}, while others found a dramatic reversal of scrambling~\cite{swingle2018resilience,zhang2019information}. We attribute this difference to the presence or absence of conservation laws.

\emph{Discussion and outlook.}---Our results lead to a number of implications.
First, we provide a new perspective on protocols which divide error-prone OTOC measurements by an independent characterization of the error~\cite{swingle2018resilience,vermersch2019probing,mi2021information,sanchez2019emergent}.
In our language, the latter is precisely the normalization, $\Norm(t) =  \langle \h{M}(t)  \, \h{M}(t) \rangle $.
To this end, these protocols will only  replicate unitary dynamics when the total error is small ($1-\Norm \approx \err \int_0^t dt \, \AvgSize \ll 1$) or when size distributions are tightly peaked, $\delta \Size \ll \Size$. 

Second, our results suggest a novel protocol for measuring operator size distributions (Fig.~\ref{fig: circuit}), which circumvents the  need to either perform exponentially many measurements~\cite{qi2019measuring} or  utilize two entangled copies of the system~\cite{schuster2021many}. 
Specifically, in order to measure the \emph{generating function} of the size distribution, $G_\Size(\mu) = \sum_\Size P(\Size) e^{-\mu \Size}$, we propose the following protocol (Fig.~\ref{fig: circuit}): (i) prepare an initial state, $\rho = (\mathbbm{1}+\h{M}) \otimes \mathbbm{1}^{\otimes N-1}/2^N$, (ii) time-evolve forward, e.g.~via a unitary operation, $U$, (iii)  apply a set of single-qubit Pauli operators, $\{ P_1, \ldots, P_N \}$, (iv) time-evolve backward via $U^\dagger$, and (v) measure $\h{M}$.
If the intervening Pauli operators are fixed, this reduces to previous schemes for measuring OTOCs~\cite{garttner2017measuring}.
However, if one randomly samples each Pauli matrix in each experimental shot, with probability $p = (1-e^{-\mu})/4$ to be $\{X,Y,Z\}$ and probability $1-3p$ to be the identity, this in effect implements a decoherence channel, $e^{-\mu \Size}$, that explicitly depends on the size superoperator. 
The fidelity to recover the initial state then gives the generating function via $\mathcal{F} = \frac{1}{2} [1+ \mathcal{N} G_\Size(\mu) ]$, where $\mathcal{N}$ can be measured by setting $\mu=0$. 

Finally, we conjecture that our framework also applies to an alternate scenario (often explored in experiments~\cite{sanchez2019emergent,sanchez2021emergent,dominguez2021decoherence,dominguez2021dynamics}), where one evolves forward  via a Hamiltonian, $H$, and backward via a perturbed Hamiltonian, $-H + \eta \, \delta H $~\cite{fn6,jalabert2001environment}.
Naively, this scenario features perturbations that are highly correlated in time and space, and thus outside the Lindbladian framework.
However, in a chaotic many-body system, one expects such correlations to quickly decay outside of some thermalization time- ($\tau_{\text{th}}$) and length-scale ($\xi_{\text{th}}$).
This assumption leads to a Fermi's golden rule~\cite{jalabert2001environment,cucchietti2003decoherence} estimate of an effective decoherence rate, $\partial_t \log(\mathcal{N}) \sim \eta^2 \tau_{\text{th}} \xi_{\text{th}} \overline{\Size}$, which scales linearly with the average operator size~\cite{suppinfo}.
Somewhat intriguingly, recent NMR experiments~\cite{dominguez2021decoherence}  precisely observe this linear scaling with operator size for $\eta \gtrsim 0.1$; this transitions to a square root scaling at smaller $\eta$ and developing a microscopic understanding of this regime remains an open question. 

Looking forward, our results also have implications for the classical simulability of open quantum systems---if operator sizes are bounded from above by a constant, $\Size_\err$, then time-evolution is in principle efficiently simulable, since the dimension of the accessible operator Hilbert space is polynomial in the system size, $\sim \! N^{\Size_\err}$.
A similar idea was recently proposed in diffusive 1D spin chains~\cite{rakovszky2020dissipation}; our results suggest that it may hold  more broadly.

\FigureCircuit

\emph{Acknowledgements}---We are grateful to Ehud Altman, Anish Kulkarni and Rahul Sahay for illuminating discussions and to Bryce Kobrin and Francisco Machado for detailed feedback on the manuscript. The numerical simulations performed in this work used the dynamite Python frontend~\cite{dynamite}, which supports a matrix-free implementation of Krylov subspace methods based on the PETSc and SLEPc packages~\cite{hernandez2005slepc}.
This work was supported by the U.S. Department of Energy, Office of Science, National Quantum Information Science Research Centers, Quantum Systems Accelerator (QSA) and by the U.S. Department of Energy, Quantum Information Science Enabled Discovery (QuantISED) for High Energy Physics (KA2401032).
T.S. acknowledges support from the National Science Foundation (QII-TAQS
program and GRFP). 

\bibliographystyle{apsrev4-1}
\bibliography{refs_size_error} 

\begin{thebibliography}{59}
\bibitem [{fn5()}]{fn5}%
  \BibitemOpen
\bibfield  {journal} {  }\href@noop {} {\bibinfo  {journal} {We derive
  Eqs.~(5,6) by differentiating $\mathcal{N}$ and $\overline{\mathcal{S}}$ with
  respect to $t$ and applying Eq.~(4)}\ }\BibitemShut {NoStop}%
\bibitem [{fn6()}]{fn6}%
  \BibitemOpen
  \href@noop {} {\bibinfo  {journal} {Note that our original use of the term
  Loschmidt echo refers to the case where one evolves forwards and backwards
  with the same Hamiltonian in the presence of noise and errors. The term
  Loschmidt echo is also used to describe the scenario here, where one evolves
  forwards and backwards with two different Hamiltonians}\ }\BibitemShut
  {NoStop}%
\end{thebibliography}

\clearpage 

\section*{Supplementary material}

# Supplemental Material: Operator Growth in Open Quantum Systems

Thomas Schuster[1] and Norman Y. Yao[1, 2]

[1]*Department of Physics, University of California, Berkeley, California 94720 USA*
[2]*Materials Science Division, Lawrence Berkeley National Laboratory, Berkeley, California 94720, USA*
(Dated: August 24, 2022)

## CONTENTS

## ADDITIONAL ANALYTICAL DETAILS AND DISCUSSION

In this section, we provide substantially more details and discussion on the approximations leading to Eq. (4) of the main text. We begin by clarifying our notion of operator growth under open-system dynamics, and outlining how this notion connects to different protocols that measure information scrambling. The remainder of the section serves to build intuition and outline the explicit steps that lead to Eq. (4) of the main text. While these steps involve a substantial number of equations, we emphasize that the physical intuition behind Eq. (4) is fully contained in the main text.

### Preliminaries

Defining operator growth under open-system dynamics involves a few subtleties not present in unitary dynamics. In the main text, we focused on the time-evolution of operators in the Heisenberg picture [Eq. (1) of the main text]:

$$\partial_t \hat{M}_H(t) = i[\hat{H}, \hat{M}_H(t)] + \varepsilon \sum_\alpha \left( \hat{L}_\alpha^\dagger \hat{M}_H(t) \hat{L}_\alpha - \frac{1}{2}\{\hat{L}_\alpha^\dagger \hat{L}_\alpha, \hat{M}_H(t)\} \right). \quad (1)$$

Heisenberg time-evolution is relevant when $\hat{M}$ is a Hermitian observable being measured in an experiment. The growth of $\hat{M}_H(t)$ signifies that the measurement outcome depends on increasingly non-local correlations in the initial state of the system. Note that Heisenberg time-evolution is unital, i.e. it maps the identity to the identity.

One can also, in principle, consider the growth of operators in the Schrodinger picture. For instance, consider preparing a system in an initial density matrix that is almost entirely mixed, $\rho = \hat{M} \otimes \mathbb{1}^{N-1}/2^N$, where $\hat{M}$ is an

initial single-qubit density matrix. The state now evolves via:

$$\partial_t \hat{M}_S(t) = -i[\hat{H}, \hat{M}_S(t)] + \varepsilon \sum_\alpha \left( \hat{L}_\alpha \hat{M}_S(t) \hat{L}_\alpha^\dagger - \frac{1}{2} \{\hat{L}_\alpha^\dagger \hat{L}_\alpha, \hat{M}_S(t)\} \right), \tag{2}$$

which differs from Heisenberg time-evolution in the sign of the Hamiltonian and the ordering of the Lindblad operators. Note that Schrodinger and Heisenberg time-evolution coincide precisely when the Lindblad operators are Hermitian, $\hat{L}_\alpha = \hat{L}_\alpha^\dagger$, in which case we have $\hat{M}_H(t; \hat{H}) = \hat{M}_S(t; -\hat{H})$. This is trivially the case for unitary dynamics, and also holds for a variety of error models of practical interest (see the subsequent subsection).

Interpreting operator growth in the Schrodinger picture outside of the case $\hat{L}_\alpha = \hat{L}_\alpha^\dagger$ is more difficult. Fundamentally, this difficulty arises because Schrodinger time-evolution is not unital, i.e. the identity operator is time-evolved non-trivially. We illustrate this point with two examples. First, take the Lindblad operators to be zero and the Hamiltonian to implement unitary scrambling dynamics such that the operator $\hat{M}_S(t)$ grows to have support across the entire system. In this case, operator growth corresponds to the formation of non-local correlations under time-evolution. Second, take the Hamiltonian to be zero and the Lindbladian operators to implement spontaneous emission on each qubit (i.e. $\hat{L}_\alpha$ are lowering operators on each qubit). At late enough times, the density matrix will approach the all zero state, $\hat{M}_S(t) \to |0\ldots0\rangle\langle0\ldots0|$. This operator is quite large (it has non-trivial support on every qubit of the system), yet its growth clearly does not represent any non-local correlations, since the system remains in a product state throughout time-evolution. For these reasons, in this work we direct our attention to operator growth under Heisenberg time-evolution.

The "correct" time-evolution is dictated by the experimental protocol under study. Let us specify to the protocol depicted in Fig. 4 of the main text, and consider the behavior of the protocol when the unitaries, $U$ and $U^\dagger$, are instead replaced by open-system dynamics under Hamiltonians, $\hat{H}$ and $-\hat{H}$, and Lindblad operators, $\hat{L}_\alpha$. Here, for simplicity we assume the Lindblad operators are the same for forwards and backwards time-evolution, although this assumption is not essential in the subsequent sections. The operator $\hat{M}$ appears in two places in the protocol: First, describing the initial single-qubit state, $|\psi_\pm\rangle\langle\psi_\pm| = (1 \pm \hat{M})/2$; and second, describing the final observable to be measured, again $(1 \pm \hat{M})/2$. Here, we take $\hat{M}$ to be a single-qubit operator, and generalize from preparing only the positive eigenstate of $\hat{M}$ (as considered in the main text) to either the positive or negative eigenstate. The expectation value of the final measurement represents the fidelity to recover the initial state. The fidelity consists of one copy each of the Schrodinger and Heisenberg time-evolved operators,

$$\begin{aligned}
\mathcal{F}_\pm &= \left\langle \frac{\hat{\mathbb{1}}_S(t; -\hat{H}) \pm \hat{M}_S(t; -\hat{H})}{2^N} \cdot e^{-\mu \mathcal{S}} \left\{ \frac{\hat{\mathbb{1}} \pm \hat{M}_H(t; \hat{H})}{2} \right\} \right\rangle \\
&= \frac{1}{2} \pm \left\langle \hat{\mathbb{1}}_S(t; -\hat{H}) \cdot e^{-\mu \mathcal{S}} \left\{ \hat{M}_H(t; \hat{H}) \right\} \right\rangle + \frac{1}{2} \left\langle \hat{M}_S(t; -\hat{H}) \cdot e^{-\mu \mathcal{S}} \left\{ \hat{M}_H(t; \hat{H}) \right\} \right\rangle,
\end{aligned} \tag{3}$$

where in the final expression we used the property, $\text{tr}(\hat{M}_S(t)) = 0$, to neglect one cross term. We can eliminate the middle term on the RHS by averaging the fidelity over the positive and negative eigenstates, $\mathcal{F} = (\mathcal{F}_+ + \mathcal{F}_-)/2$. We have:

$$\mathcal{F} = \frac{1}{2} + \frac{1}{2} \left\langle \hat{M}_S(t; -\hat{H}) \cdot e^{-\mu \mathcal{S}} \left\{ \hat{M}_H(t; \hat{H}) \right\} \right\rangle, \tag{4}$$

which is the result quoted in the main text, upon the substitution $\hat{M}(t) \equiv \hat{M}_S(t; -\hat{H}) = \hat{M}_H(t; \hat{H})$ for Hermitian Lindblad operators.

We briefly comment on how open-system dynamics manifest in other protocols for measuring information scrambling. The above analysis applies to a large class of protocols that involve evolving once via forward time-evolution and once via backwards time-evolution [1–6]. In contrast, protocols that measure scrambling via exclusively forward time-evolution [7, 8] will depend instead on inner products of two copies of the Heisenberg time-evolved operator (although we note that such protocols must in some cases be exponentially inefficient [9]). Finally, in teleportation-based protocols [10–14], the operator of interest enters neither a measured observable nor as an initial state, but rather as a quantum operation applied to one half of EPR pairs. In this case, one should properly consider the time-evolution of the *quantum channel*, $\mathcal{M}\{\cdot\} \equiv \hat{M}(\cdot)\hat{M}^\dagger$, which displays qualitatively different behavior from the Heisenberg time-evolution analyzed in this work [14].

## Examples of Lindblad operators

We can build intuition for the effect of Lindbladian time-evolution on operator growth in the Heisenberg and Schrodinger pictures by analyzing a few simple error models. Let us begin with the simplest case, of isotropic decoherence on each qubit. In this case, we found in the main text that the action of the Lindbladian was precisely given by the size superoperator:

$$-\frac{1}{4}\sum_i \sum_P (P_i \hat{M}(t) P_i^\dagger - \hat{M}(t)) = \mathcal{S}\{\hat{M}(t)\} = \# \text{ of } X, Y, Z \text{ in } \hat{M}(t), \tag{5}$$

where the RHS is intended to indicate that the size superoperator multiplies each Pauli string, $R$, of the operator, $\hat{M}(t) = \sum_R c_R \hat{M}(t)$, by its number of non-identity components (i.e. the number of $X$-, $Y$-, or $Z$-components). Here the Pauli matrices run over $P \in \{\mathbb{1}, X, Y, Z\}$ on each qubit.

Different error models do not exactly measure the operator size. For instance, dephasing noise gives rise to the following superoperator:

$$-\frac{1}{2}\sum_i (Z_i \hat{M}(t) Z_i^\dagger - \hat{M}(t)) = \mathcal{S}_Z\{\hat{M}(t)\} = \# \text{ of } X, Y \text{ in } \hat{M}(t), \tag{6}$$

while depolarizing noise gives rise to:

$$-\frac{1}{4}\sum_i (X_i \hat{M}(t) X_i^\dagger + Y_i \hat{M}(t) Y_i^\dagger - 2\hat{M}(t)) = \left[\mathcal{S} - \frac{1}{2}\mathcal{S}_Z\right]\{\hat{M}(t)\} = \left(\# \text{ of } Z \text{ in } \hat{M}(t)\right) + \frac{1}{2}\left(\# \text{ of } X, Y \text{ in } \hat{M}(t)\right). \tag{7}$$

In the main text and subsequent section, we argue that these superoperators display similar phenomenology as the operator size when applied to chaotic many-body time-evolved operators. From the definitions above, we see that this is a reasonable assumption if the high-size components of time-evolved operators fluctuate randomly between the local choice of basis (i.e. in these examples, between single-qubit $X$-, $Y$-, and $Z$-components).

Non-Hermitian errors require more careful consideration than Hermitian errors, since their time-evolution differs between the Heisenberg and Schrodinger pictures. For instance, consider spontaneous emission on each qubit, i.e. Lindblad operators equal to the lowering operator, $\hat{\ell}_i = (\hat{X}_i - i\hat{Y}_i)/2$. We denote the raising operator, $\hat{r}_i = \hat{\ell}_i^\dagger$, the projector to the up eigenstate, $\hat{U}_i = \hat{r}_i \hat{\ell}_i = (\mathbb{1}_i + \hat{Z}_i)/2$, and the projector to the down eigenstate, $\hat{D}_i = \hat{\ell}_i \hat{r}_i = \mathbb{1}_i - \hat{U}_i$. In the Heisenberg picture, operators time-evolve via:

$$\partial_t \hat{M}_H(t) = i[\hat{H}, \hat{M}_H(t)] + \varepsilon \sum_i \left(\hat{r}_i \hat{M}_H(t) \hat{\ell}_i - \frac{1}{2}\{\hat{U}_i, \hat{M}_H(t)\}\right). \tag{8}$$

Let us focus on the effect of spontaneous emission on a single component, $R_i$, of a single Pauli string, $\hat{R} = \bigotimes_i R_i$. We have:

$$-\hat{r}_i \hat{R}_i \hat{\ell}_i + \frac{1}{2}\{\hat{U}_i, \hat{R}_i\} = \begin{cases} 0, & \hat{R}_i = \mathbb{1}_i \\ \hat{X}_i/2, & \hat{R}_i = \hat{X}_i \\ \hat{Y}_i/2, & \hat{R}_i = \hat{Y}_i \\ 2\hat{U}_i, & \hat{R}_i = \hat{Z}_i \end{cases}. \tag{9}$$

We see that the error cannot increase the size of a Pauli string, since each identity component is mapped to itself under Heisenberg time-evolution. Summing over all qubits, we have:

$$-\sum_i \left(\hat{r}_i \hat{R} \hat{\ell}_i - \frac{1}{2}\{\hat{U}_i, \hat{R}\}\right) = \left[\mathcal{S} - \frac{1}{2}\mathcal{S}_Z\right]\{\hat{R}\} - \sum_i (\delta_{R_i = Z_i}) \, \hat{\mathbb{1}}_i \otimes \hat{R}_{\neq i}. \tag{10}$$

The effect of spontaneous emission in the Heisenberg picture is thus a combination of depolarizing noise (left term) and a superoperator that generates Pauli strings that differ from $\hat{R}$ by replacement of a single $Z$-component with the identity (right term). We do not expect the latter effect to substantially alter operator spreading, since the newly generated strings contain similar support to the original string. Indeed, in the following subsection we find that their effect is suppressed by a factor of $\sim \varepsilon/J$, where $J$ is the local Hamiltonian interaction strength.

As noted in the previous subsection, the time-evolution of operators in the Schrodinger picture can be drastically different. Again considering spontaneous emission, we have:

$$\partial_t \hat{M}_S(t) = -i[\hat{H}, \hat{M}_S(t)] + \varepsilon \sum_i \left( \hat{\ell}_i \hat{M}_S(t) \hat{r}_i - \frac{1}{2}\{\hat{U}_i, \hat{M}_S(t)\} \right), \tag{11}$$

where the Lindblad operators now act as,

$$-\hat{\ell}_i \hat{R}_i \hat{r}_i + \frac{1}{2}\{\hat{U}_i, \hat{R}_i\} = \begin{cases} \hat{Z}_i, & \hat{R}_i = \mathbb{1}_i \\ \hat{X}_i/2, & \hat{R}_i = \hat{X}_i \\ \hat{Y}_i/2, & \hat{R}_i = \hat{Y}_i \\ \hat{Z}_i, & \hat{R}_i = \hat{Z}_i \end{cases}. \tag{12}$$

Summing over all qubits, we now have:

$$-\sum_i \left( \hat{r}_i \hat{R} \hat{\ell}_i - \frac{1}{2}\{\hat{U}_i, \hat{R}\} \right) = \left[ \mathcal{S} - \frac{1}{2}\mathcal{S}_Z \right]\{\hat{R}\} - \sum_i \left( \delta_{R_i = \mathbb{1}_i} \right) \hat{Z}_i \otimes \hat{R}_{\neq i}. \tag{13}$$

The effect of spontaneous emission in the Schrodinger picture is thus a combination of depolarizing noise (left term) and a superoperator that generates Pauli strings that differ from $\hat{R}$ by a single replacement of an *identity* component with a Pauli-$Z$ operator (right term). As in the previous subsection, we see that non-Hermitian Lindblad operators cause Schrodinger time-evolved operators to immediately gain support on all qubits, irrespective of the locality of the dynamics.

### Approximations leading to Eq. (4) of the main text

In the main text, we provided intuitive arguments that open-system operator growth is governed by an effective Lindblad equation involving the size superoperator [Eq. (4) of the main text], in the case of chaotic dynamics and small error rates. In this subsection, we specify the precise approximations which lead to Eq. (4) of the main text.

We begin by formally solving the Heisenberg and Schrodinger time-evolution equations as a Taylor series in the error rate. We have:

$$\hat{M}_H(t) = \sum_{n=0}^\infty \varepsilon^n \sum_{\{\alpha_i\}} \int_{t_1 \leq \ldots \leq t_n} dt_1 \ldots dt_n e^{-iH(t-t_n)} \mathcal{L}_{\alpha_n}\{\ldots \mathcal{L}_{\alpha_1}\{e^{-iHt_1} M e^{iHt_1}\}\ldots\} e^{iH(t-t_n)}, \tag{14}$$

and

$$\hat{M}_S(t) = \sum_{n=0}^\infty \varepsilon^n \sum_{\{\alpha_i\}} \int_{t_1 \leq \ldots \leq t_n} dt_1 \ldots dt_n e^{-iH(t-t_n)} \mathcal{L}^\dagger_{\alpha_n}\{\ldots \mathcal{L}^\dagger_{\alpha_1}\{e^{-iHt_1} M e^{iHt_1}\}\ldots\} e^{iH(t-t_n)}, \tag{15}$$

where for each Lindblad operator, $\hat{L}_\alpha$, we associate a superoperator, $\mathcal{L}_\alpha\{\cdot\} = \hat{L}^\dagger_\alpha(\cdot)\hat{L}_\alpha - \frac{1}{2}\{\hat{L}^\dagger_\alpha \hat{L}_\alpha, \cdot\}$, and its adjoint, $\mathcal{L}^\dagger_\alpha\{\cdot\} = \hat{L}_\alpha(\cdot)\hat{L}^\dagger_\alpha - \frac{1}{2}\{\hat{L}^\dagger_\alpha \hat{L}_\alpha, \cdot\}$. Inserting these solutions into Eq. (4), we find that the protocol in Fig. 4 of the main text measures the following quantity:

$$\left\langle \hat{M}_S(t; -\hat{H}) e^{-\mu \mathcal{S}} \left\{ \hat{M}_H(t; \hat{H}) \right\} \right\rangle$$
$$= \sum_{m,n=0}^\infty \varepsilon^{n+m} \sum_{\{\alpha_i\}} \sum_{\{\beta_i\}} \int_{s_1 \leq \ldots \leq s_m} ds_1 \ldots ds_m \int_{t_1 \leq \ldots \leq t_n} dt_1 \ldots dt_n \times$$
$$\left\langle e^{-iH(t-s_m)} \mathcal{L}^\dagger_{\beta_m}\{\ldots \mathcal{L}^\dagger_{\beta_1}\{e^{-iHs_1} M e^{iHs_1}\}\ldots\} e^{iH(t-s_n)} e^{-\mu \mathcal{S}}\{e^{-iH(t-t_n)} \mathcal{L}_{\alpha_n}\{\ldots \mathcal{L}_{\alpha_1}\{e^{-iHt_1} M e^{iHt_1}\}\ldots\} e^{iH(t-t_n)}\} \right\rangle. \tag{16}$$

Each expectation value in the summation is a multi-point OTOC involving $\hat{M}$, the Lindblad operators, and the Pauli operators used to implement the quantum channel, $e^{-\mu \mathcal{S}}$.

We apply two approximations to this expression: (i) First, we assume that the precise form of operators entering the OTOC is only relevant for operators that occur at approximately the same "time slice", i.e. their times are separated by less than the inverse local interaction strength, $\sim 1/J$. (ii) Second, we assume that the only OTOCs that are non-zero are those for which all operators in every time-slice product to the identity, i.e. "off-diagonal" OTOCs such as $\langle \hat{Z}_i(t) \hat{X}_j(0) \hat{Z}_i(t) \hat{Y}_j(0) \rangle$ are zero. Properties (i) and (ii) are exact within large-$N$ systems and random matrix theory [15], and in random unitary circuits for single-qubit errors at leading order in the error rate. In the latter, condition (i) follows because the local basis is randomized with each time step, while condition (ii) follows because the sign of off-diagonal OTOCs can be flipped via single-qubit rotations (e.g. conjugation by the Pauli operator $X_j$ at time zero in the aforementioned example), which thus average the OTOC to zero. Corrections arise when the same Lindblad superoperator is applied in both forwards and backwards time-evolution, which occurs with a probability suppressed by $\varepsilon^2$. The application of these approximations to chaotic many-body systems is a conjecture, motivated by the intuition presented in the main text.

We now apply approximations (i) and (ii) to Eq. (16). In particular, we show that approximation (ii) implies that operator growth under non-Hermitian Lindblad operators is equal to that under a particular set of Hermitian Lindblad operators, to leading order in $\varepsilon/J$. Since Schrodinger and Heisenberg time-evolution are equivalent for Hermitian Lindblad operators, this verifies that the operator size distribution measured via Eq. (4) is well-defined (i.e. it contains no negative elements). From this result, approximation (i) leads to Eq. (4) of the main text.

To establish the first simplification, we define Hermitian operators, $\hat{L}_\alpha^+ = (\hat{L}_\alpha + \hat{L}_\alpha^\dagger)/2$ and $\hat{L}_\alpha^- = i(\hat{L}_\alpha - \hat{L}_\alpha^\dagger)/2$. Note that the inner product between the two Hermitian operators, $\langle \hat{L}_+ \hat{L}_- \rangle = \frac{1}{2}[\langle \hat{L}^\dagger \hat{L}^\dagger \rangle - \langle \hat{L} \hat{L} \rangle]$, can be set to zero by multiplying $\hat{L}$ by an overall phase. The Lindblad superoperator can be expressed as:

$$\begin{aligned}
\mathcal{L}_\alpha \{\cdot\} &= \hat{L}_\alpha (\cdot) \hat{L}_\alpha^\dagger - \frac{1}{2} \{\hat{L}_\alpha^\dagger \hat{L}_\alpha, (\cdot)\} \\
&= \mathcal{L}_\alpha^+ \{\cdot\} + \mathcal{L}_\alpha^- \{\cdot\} - i \hat{L}_\alpha^- (\cdot) \hat{L}_\alpha^+ + i \hat{L}_\alpha^+ (\cdot) \hat{L}_\alpha^- - \frac{i}{2} \{[\hat{L}_\alpha^-, \hat{L}_\alpha^+], (\cdot)\} \\
\mathcal{L}_\alpha^\dagger \{\cdot\} &= \hat{L}_\alpha^\dagger (\cdot) \hat{L}_\alpha - \frac{1}{2} \{\hat{L}_\alpha^\dagger \hat{L}_\alpha, (\cdot)\} \\
&= \mathcal{L}_\alpha^+ \{\cdot\} + \mathcal{L}_\alpha^- \{\cdot\} + i \hat{L}_\alpha^- (\cdot) \hat{L}_\alpha^+ - i \hat{L}_\alpha^+ (\cdot) \hat{L}_\alpha^- - \frac{i}{2} \{[\hat{L}_\alpha^-, \hat{L}_\alpha^+], (\cdot)\}
\end{aligned} \quad (17)$$

where $\mathcal{L}_\alpha^\pm \{\cdot\}$ are the analogous Lindblad superoperators for $\hat{L}_\alpha^\pm$.

Now, note that the typical separation in time of spatially and temporally nearby Lindblad operators in Eq. (16) is given by the inverse local error rate, $\sim 1/\varepsilon$. For small error rates, this is much longer than the thermalization time, $\sim 1/J$; thus we can assume that each Lindblad superoperator in Eq. (16) occurs at a different "time-slice" for the purposes of approximations (i), (ii). Applying Eq. (17), we see that each time-slice contains at most two of a given Lindblad operator: either two copies of $\hat{L}_\alpha^+$ (from $\mathcal{L}_\alpha^+ \{\cdot\}$), two copies of $\hat{L}_\alpha^-$ (from $\mathcal{L}_\alpha^- \{\cdot\}$), or one copy each of $\hat{L}_\alpha^+$ and $\hat{L}_\alpha^-$ (from the latter four terms). However, the latter terms are precisely those that vanish under approximation (ii). Therefore, to leading order in $\varepsilon/J$, we find that time-evolution is effectively governed by the Hermitian Lindblad operators, $\hat{L}_\alpha^\pm$. Concretely, we can replace $\hat{M}_H(t; \hat{H}) \to \hat{M}(t)$ and $\hat{M}_S(t; -\hat{H}) \to \hat{M}(t)$ in Eq. (16), where the operator, $\hat{M}(t)$, evolves as:

$$\partial_t \hat{M}(t) = i[\hat{H}, \hat{M}(t)] + \varepsilon \sum_\alpha \left( \mathcal{L}_\alpha^+ \{\hat{M}(t)\} + \mathcal{L}_\alpha^- \{\hat{M}(t)\} \right). \quad (18)$$

For example, in the case of spontaneous emission, with non-Hermitian Lindblad operator $\hat{L}_i = \hat{\ell}_i$, we have Hermitian Lindblad operators, $\hat{L}_i^+ = X$ and $\hat{L}_i^- = Y$, which represent an effective depolarizing channel. Intuitively, this occurs because the Pauli strings in the final right term of Eqs. (10,13) are effectively uncorrelated with those of the original operator after time $\sim 1/J$, and thus only contribute at higher order in $\varepsilon/J$.

Applying approximation (i) to Eq. (18) leads us to Eq. (4) in the main text. Specifically, we replace each Lindblad superoperator, $\mathcal{L}_\alpha$, with a sum of Pauli Lindblad superoperators, $\mathcal{L}_{P_i}$, acting on the same qubit:

$$\partial_t \hat{M}(t) \approx i[\hat{H}, \hat{M}(t)] + \tilde{\varepsilon} \sum_i \sum_{P_i} \mathcal{L}_{P_i} \{\hat{M}(t)\} = i[\hat{H}, \hat{M}(t)] + 4\tilde{\varepsilon} \mathcal{S}\{\hat{M}(t)\}. \quad (19)$$

To be precise, we can adjust the local error rates so that the normalization of the Lindblad operators is unchanged, $\varepsilon \langle \hat{L}_\alpha^\dagger \hat{L}_\alpha \rangle = \tilde{\varepsilon} \sum_{P_i \neq \mathbb{1}_i} \langle \hat{P}_i \hat{P}_i \rangle = 3\tilde{\varepsilon}$. In the remainder of the work, we replace the notation $\tilde{\varepsilon}$ with $\varepsilon$ for simplicity. If

$\mathcal{L}_\alpha$ has support on multiple qubits but the qubits are nearby each other in space, we expect this approximation to apply when $i$ is chosen as any one of the qubits, because the support of $\hat{M}(t)$ will be highly correlated between nearby qubits in a locally interacting system. If $\mathcal{L}_\alpha$ has support on multiple qubits in an all-to-all coupled system, we expect the approximation to apply when $i$ is summed over all qubits, since the support of $\hat{M}(t)$ will be nearly entirely uncorrelated between qubits.

Before concluding, we mention a few notable violations of the approximations (i) and (ii), and show that they nonetheless have a subleading effect. First, note that systems with local conserved quantities violate both approximations. Regarding approximation (i), Lindblad operators that overlap vs. do not overlap the conserved quantity will spread differently; for approximation (ii), certain terms with zero trace in each time-slice are non-zero, for example the two-point function, $\langle \hat{Z}_i(t)\hat{Z}_j(0)\rangle$. Nonetheless, we expect both of these violations to give only subleading contribution (suppressed by $1/\sqrt{t}$), since the majority of a time-evolved operator's support is contained in chaotic large-size Pauli strings.

Second, a distinct argument shows that the approximation (ii) is violated in Hamiltonian systems. Consider the time-derivative of the OTOC under unitary dynamics:

$$\partial_t \langle \hat{M}(t)\hat{Q}\hat{M}(t)\hat{Q}\rangle = i\langle \hat{M}(t)[\hat{Q},\hat{H}]\hat{M}(t)\hat{Q}\rangle. \tag{20}$$

The right side is an "off-diagonal" OTOC since the trace of all operators at time zero vanishes, yet it is non-zero whenever the OTOC on the left side changes in time. We now show that this violation gives only a subleading contribution to Eq. (4) in two examples. In all-to-all interacting systems, the right side of Eq. (20) contains a sum of at least $\sim N$ Hamiltonian terms, which leads to a $\sim 1/N$ suppression in the value of each off-diagonal OTOC. Meanwhile, in a 1D system, the left side OTOC (before differentiation) is zero when the operator $Q$ is within the light cone of $\hat{M}(t)$, and unity when it is outside. The time-derivative of the OTOC is therefore non-zero only over a small region of width $\sim c\sqrt{t}$, with a magnitude $\sim v_B/c\sqrt{t}$. The total contribution of "off-diagonal" OTOCs to Eq. (4) therefore scales as $\varepsilon(c\sqrt{t})(v_B/c\sqrt{t}) \sim \varepsilon \partial_t \mathcal{S}/J \sim \varepsilon$, which is much less than the contribution of "diagonal" OTOCs, $\sim \varepsilon \mathcal{S}$.

### Derivation of Eqs. (5), (6) of the main text

In this section, we derive the Loschmidt echo decay equation [Eq. (5) of the main text] and the size growth equation [Eq. (6) of the main text]. Both equations are obtained directly by applying Eq. (4) of the main text to the quantities of interest. We begin with the Loschmidt echo fidelity, $\mathcal{N} = \langle \hat{M}(t)\hat{M}(t)\rangle$. The time-derivative is:

$$\begin{aligned}\partial_t \mathcal{N}(t) &= 2\langle \partial_t \hat{M}(t)\hat{M}(t)\rangle \\ &= -2\varepsilon \langle \mathcal{S}\{\hat{M}(t)\}\hat{M}(t)\rangle \\ &= -2\varepsilon \overline{\mathcal{S}}(t)\mathcal{N}(t).\end{aligned} \tag{21}$$

using the definition of the average operator size, $\overline{\mathcal{S}}(t) = \langle \mathcal{S}\{\hat{M}(t)\}\hat{M}(t)\rangle/\langle \hat{M}(t)\hat{M}(t)\rangle$. Taking the time-derivative of the average operator size gives:

$$\begin{aligned}\partial_t \overline{\mathcal{S}}(t) &= \frac{\partial_t \langle \hat{M}(t)\mathcal{S}\{\hat{M}(t)\}\rangle}{\langle \hat{M}(t)\hat{M}(t)\rangle} - \frac{\langle \hat{M}(t)\mathcal{S}\{\hat{M}(t)\}\rangle \cdot \partial_t \langle \hat{M}(t)\hat{M}(t)\rangle}{\langle \hat{M}(t)\hat{M}(t)\rangle^2} \\ &= 2i\left(\frac{\langle [\hat{H},\hat{M}(t)]\mathcal{S}\{\hat{M}(t)\}\rangle}{\langle \hat{M}(t)\hat{M}(t)\rangle}\right) - 2\varepsilon\left(\frac{\langle \hat{M}(t)\mathcal{S}^2\{\hat{M}(t)\}\rangle}{\langle \hat{M}(t)\hat{M}(t)\rangle}\right) + 2\varepsilon\left(\frac{\langle \hat{M}(t)\mathcal{S}\{\hat{M}(t)\}\rangle^2}{\langle \hat{M}(t)\hat{M}(t)\rangle^2}\right) \\ &= 2i\left(\frac{\langle [\hat{H},\hat{M}(t)]\mathcal{S}\{\hat{M}(t)\}\rangle}{\langle \hat{M}(t)\hat{M}(t)\rangle}\right) - 2\varepsilon \delta\mathcal{S}(t)^2,\end{aligned} \tag{22}$$

where the size width is defined as $\delta\mathcal{S}(t)^2 = \overline{\mathcal{S}^2}(t) - \overline{\mathcal{S}}(t)^2$. The first term corresponds to size growth via the unitary dynamics generated by $\hat{H}$.

## Unequal Hamiltonian formulation of the Loschmidt echo

In this section, we address the alternate Loschmidt echo scenario discussed in the main text, in which time-evolution is unitary but is governed by unequal Hamiltonians for forwards and backwards time-evolution. Similar to the previous subsections, we outline the precise approximations which lead to a Loschmidt echo fidelity described by Eq. (4) of the main text. In what follows, we begin by Taylor expanding the fidelity in the difference between the two Hamiltonians, in which case each term is again a sum of various OTOCs. We then apply a single assumption: that these correlation functions relax quickly outside of some correlation length, $\xi_{\text{th}}$, and correlation time, $\tau_{\text{th}}$. We expect these assumptions to hold for chaotic many-body systems when the perturbation $\delta H$ does not overlap any conserved quantities (e.g. the Hamiltonian, $H$). In principle, we expect modifications to our predictions when the perturbation does overlap a conserved quantity. In Fig. 3 of the main text we explore this scenario numerically for small system sizes, and observe qualitatively similar behavior to that expected from Lindbladian dynamics.

The fidelity of the many-body Loschmidt echo is given by the correlation function:

$$\mathcal{F}_\eta(t) = \langle e^{-iHt} M e^{iHt} e^{-i(H+\eta\,\delta H)t} M e^{i(H+\eta\,\delta H)t}\rangle. \tag{23}$$

We begin our derivation by Taylor expanding in the perturbation strength, $\eta$:

$$\mathcal{F}_\eta(t) = \sum_n (i\eta)^n \mathcal{C}_n(t) \tag{24}$$

where the $n^{\text{th}}$ coefficient is given by:

$$\mathcal{C}_n(t) = \sum_{\text{bipart}} (-1)^{n_L} \int dt_1^L \ldots dt_{n_L}^L \int dt_1^R \ldots dt_{n_R}^R \times$$
$$\text{tr}\left(e^{-iHt} M e^{iHt} e^{-iH(t-t_{n_L}^L)} \delta H \ldots e^{-iH(t_2^L - t_1^L)} \delta H e^{-iHt_1^L} M e^{iHt_1^R} \delta H e^{iH(t_2^R - t_1^R)} \ldots \delta H e^{iH(t-t_{n_R}^R)}\right) \tag{25}$$

Here the summation occurs over all bipartitions, $(n_L, n_R)$, with $n_L + n_R = n$ and $n_L, n_R \geq 0$ (since each $\eta$-derivative can hit either $e^{-i(H+\eta\,\delta H)t}$ on the left, or $e^{i(H+\eta\,\delta H)t}$ on the right). The time-ordering is $t_1^L \leq t_2^L \leq \ldots \leq t_{n_L}^L$ and $t_1^R \leq t_2^R \leq \ldots \leq t_{n_R}^R$. Each term in the above equation is an OTOC,

$$\begin{aligned}\langle e^{-iHt} M e^{iHt} e^{-iH(t-t_{n_L}^L)} \delta H \ldots e^{-iH(t_2^L-t_1^L)} \delta H e^{-iHt_1^L} M e^{iHt_1^R} \delta H e^{iH(t_2^R-t_1^R)} \ldots \delta H e^{iH(t-t_{n_R}^R)}\rangle \\ = \langle \hat{M}(t) \delta H(t-t_{n_L}^L) \ldots \delta H(t-t_1^L) \hat{M}(t) \delta H(t-t_1^R) \ldots \delta H(t-t_n^R)\rangle\end{aligned} \tag{26}$$

To proceed farther, we approximate that each OTOC is zero unless its set of times, $\{t_1^L, \ldots t_{n_L}^L, t_1^R, \ldots, t_{n_R}^R\}$, can be decomposed into pairs of equal times (note that this is equivalent to approximation (ii) in the previous subsection). Note that this immediately eliminates all $\mathcal{C}_n(t)$ with $n$ odd. These pairs can occur either between times on the same side (e.g. $t_1^L = t_2^L$ or $t_1^R = t_2^R$), or on opposite sides (e.g. $t_1^L = t_1^R$). Summing these four possibilities for a fixed set of times, we find that each pair enters in the form:

$$\mathcal{L}_{\delta H}\{(\ldots)\hat{M}(\ldots)\} = \delta H \delta H (\ldots)\hat{M}(\ldots) + (\ldots)\hat{M}(\ldots)\delta H \delta H - 2\delta H(\ldots)\hat{M}(\ldots)\delta H, \tag{27}$$

where $\mathcal{L}_{\delta H}$ is a superoperator and the ellipses represent intermediate time-evolution. The sum over bipartitions in Eq. (25) can thus be re-expressed as:

$$\mathcal{C}_{2n}(t) \approx \tau_{\text{th}}^n \int dt_1 \ldots dt_n \langle e^{-iHt} M e^{iHt} e^{-iH(t-t_n)} \mathcal{L}_{\delta H}\{\ldots e^{-iH(t_2-t_1)} \mathcal{L}_{\delta H}\{e^{-iHt_1} M e^{iHt_1}\} e^{iH(t_2-t_1)} \ldots\} e^{iH(t-t_n)}\rangle. \tag{28}$$

This can be interpreted as the connected part of the OTOC between $M$ and $n$ copies of $\delta H$.

We have not yet assumed anything about the spatial correlations of $\delta H$. However, we can already connect our expression for the Loschmidt echo fidelity to an effective Lindblad equation. Specifically, applying the above results, we can write:

$$\mathcal{F}_p(t) \approx \sum_n (-p^2)^n \mathcal{C}_{2n}(t) = \langle M_t \tilde{M}_t \rangle \tag{29}$$

with the operator $\tilde{M}_t$ defined as:

$$\tilde{M}_t \equiv \sum_n (-\tau_{\text{th}} p^2)^n \int dt_1 \ldots dt_n e^{-iH(t-t_n)} \mathcal{L}_{\delta H}\{\ldots e^{-iH(t_2-t_1)} \mathcal{L}_{\delta H}\{e^{-iHt_1} M e^{iHt_1}\} e^{iH(t_2-t_1)} \ldots\} e^{iH(t-t_n)}. \quad (30)$$

Taking a time-derivative (note that the time-derivative hits three locations: the first and final applications of $e^{\pm iH(t-t_n)}$, and the upper limit of the integral over $0 \leq t_n \leq t$), we have:

$$\begin{aligned}
\partial_t \tilde{M}_t &= -i[H, \tilde{M}_t] + \sum_n (-p^2)^n \int dt_1 \ldots dt_{n-1} \mathcal{L}_{\delta H}\{\ldots e^{-iH(t_2-t_1)} \mathcal{L}_{\delta H}\{e^{-iHt_1} M e^{iHt_1}\} e^{iH(t_2-t_1)} \ldots\} \\
&= -i[H, \tilde{M}_t] - p^2 \mathcal{L}_{\delta H}\{\tilde{M}_t\}. \\
&= -i[H, \tilde{M}_t] - 2\tau_{\text{th}} p^2 \left[\frac{1}{2} \delta H^2 \tilde{M}_t + \frac{1}{2} \tilde{M}_t \delta H^2 - \delta H \tilde{M}_t \delta H\right].
\end{aligned} \quad (31)$$

The operator is thus the solution to an effective Lindblad equation with a single Lindblad operator, $\hat{L} = \delta H$, and error rate, $\varepsilon = 2\tau_{\text{th}} p^2$. Note that we could have symmetrized the above equation by re-defining the original Hamiltonians so that $H_1 = H - \eta \, \delta H/2$ and $H_2 = H + \eta \, \delta H/2$.

To proceed further, we assume that the perturbation is a sum of local operators, $\delta H = \sum_i \delta H_i$, and that OTOCs involving a pair of local operators at the same fixed time are zero unless the operators are spatially nearby one another. Our notion of 'local' and 'nearby' can be clarified. For locally interacting systems, we have in mind that each $\delta H_i$ is contained within a finite spatial region, e.g. a ball of finite radius. We then assume that the relevant OTOCs are zero unless the pair of local terms within each fixed-time application of $\mathcal{L}_{\delta H}$ are within distance $\xi_{\text{th}}$ of each other—there are $\sim \xi_{\text{th}}^d$ such pairs in dimension $d$. In all-to-all coupled systems, we have in mind that $\delta H_i$ acts on only one of the all-to-all coupled degrees of freedom (e.g. a single qubit in an all-to-all coupled spin system, or a single Majorana fermion in SYK). The OTOCs between pairs of different $\delta H_i$ and $\delta H_j$ are then zero whenever $i \neq j$. For concreteness, we focus on local systems in what follows.

Applying this assumption to the effective Lindblad equation, Eq. (31), immediately yields a new effective Lindblad equation:

$$\partial_t \tilde{M}_t = -i[H, \tilde{M}_t] - 2\tau_{\text{th}} \xi_{\text{th}} p^2 \sum_i \left[\frac{1}{2} \delta H_i^2 \tilde{M}_t + \frac{1}{2} \tilde{M}_t \delta H_i^2 - \delta H_i \tilde{M}_t \delta H_i\right], \quad (32)$$

which involves only local Lindblad operators, $\hat{L}_i = \delta H_i$, each with error rate, $2\tau_{\text{th}} \xi_{\text{th}} p^2$. The approximation (i) in the previous subsection then leads to Eq. (4) of the main text.

## INFORMATION-THEORETIC INTERPRETATION OF THE LOSCHMIDT ECHO

In this section, we show that the Loschmidt echo with respect to a local operator $\hat{M}$ is related to a Renyi-2 mutual information, between the many-body input state of Lindbladian time-evolution and the subsystem of the output state that $\hat{M}$ act upon. Specifically, this relation holds for the average Loschmidt echo fidelity over a complete basis of operators $\hat{M}$ on the subsystem of interest.

To define the relevant subsystems, we consider a setup where our many-body system $S$ is initially maximally entangled with a reference system $E$. The state then undergoes Lindbladian time-evolution, i.e. we apply the quantum channel, $\mathcal{E} = e^{\mathcal{L}t}$, where $\mathcal{L}$ is the Lindbladian superoperator on system $S$. We are interested in the Renyi-2 mutual information between $E$ and a subsystem $A$ of $S$.

To compute the mutual information, we utilize a tensor network-like diagrammatic notation introduced in Ref. [10] (see Ref. [14] for a more extensive introduction). We represent the density matrix of the entire system, $SE$, after

Lindbladian time-evolution via the following diagram:

$$\rho_{BAE} = \boxed{\mathcal{E}} \quad (33)$$

Here system $S$ is represented via three pairs of tensor indices on the left side (decomposed into subsystem $A$ and its complement, $B$), and system $E$ similarly on the right. We denote the number of qubits in $S$ as $N$ and its Hilbert space dimension as $d = 2^N$, and similar for subsystem $A$, with $n_A$ qubits and dimension $d_A = 2^{n_A}$. For each subsystem, the tensor legs terminating downward represent the input to the density matrix (i.e. the ket indices), while the legs terminating upward represent the output (i.e. the bra indices). The middle horizontal lines represent initial EPR pairs between subsystems A, B and the reference system E. Each dot represents a factor of the inverse square root dimension of the tensor leg it is located on, to normalize the EPR pairs. The quantum channel $\mathcal{E}$ acts on system $S$, i.e. the left half of the EPR pairs.

The Renyi-2 mutual information of interest is defined as:

$$I^{(2)}(A, E) = S^{(2)}(E) + S^{(2)}(A) - S^{(2)}(AE) = N + n_A - S^{(2)}(AE), \quad (34)$$

where we use the fact $E$ and $A$ are in maximally mixed states. Here, the Renyi-2 entropy is defined as:

$$S^{(2)}(AE) = -\log_2\left[\text{tr}(\rho_{AE}^2)\right], \quad (35)$$

with $\rho_{AE} = \text{tr}_B(\rho_{BAE})$. In the diagrammatic notation, the quantity $\text{tr}(\rho_{AE}^2)$ can be expressed as:

$$\text{tr}(\rho_{AE}^2) = \quad = \frac{1}{d_A} \sum_{P_A} \quad = \frac{1}{d_A} \sum_{P_A} \quad (36)$$

The leftmost diagram follows straightforwardly from Eq. (33) by tracing over subsystem $B$, squaring, and then tracing over $A$ and $E$. In the first equality, we use the identity, $\text{tr}_A(\cdot) = \frac{1}{d_A} \sum_{P_A} \hat{P}_A(\cdot)\hat{P}_A^\dagger$, where $\hat{P}_A$ form a complete basis for operators on subsystem $A$ (e.g. the Pauli operators), to re-arrange the tensor indices. In the second equality, we complex conjugate the entire diagram and apply the definition of the adjoint in order to "slide" the channel $\mathcal{E}$ across the EPR pairs (note that $\text{tr}(\rho_{AE}^2)$ is real and so is invariant under complex conjugation; we also re-index the summation $\hat{P}_A \to \hat{P}_A^*$ for convenience). This can be derived explicitly by decomposing $\mathcal{E}$ in terms of Kraus operators, $\mathcal{E}(\cdot) = \sum_i \hat{E}_i(\cdot)\hat{E}_i^\dagger$, and utilizing the identity [14], $\hat{E}_{i,\text{left}}^T |\text{EPR}\rangle = \hat{E}_{i,\text{right}} |\text{EPR}\rangle$, for each Kraus operator. Now, we note that this diagram contains precisely the Heisenberg time-evolution of the operators, $\hat{P}_A(t) = \mathcal{E}^\dagger\{\hat{P}_A\}$. Contracting the tensor indices and keeping track of the various Hilbert space dimensions, we thus

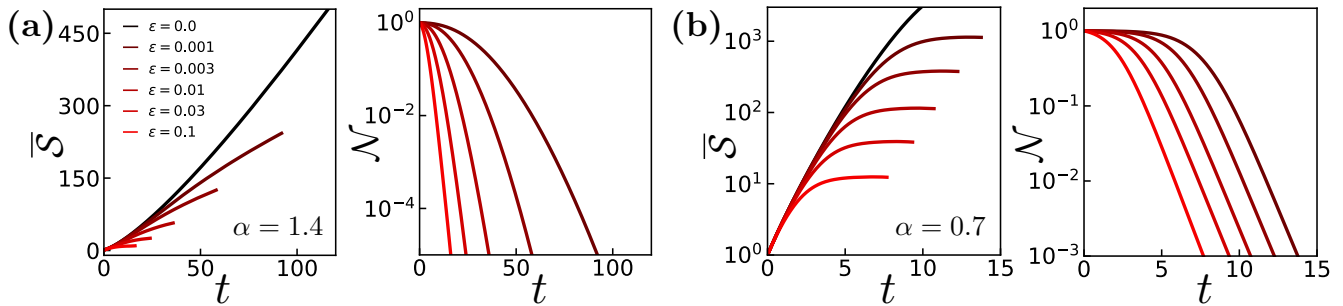

FIG. 1. Operator growth in long-range interacting random unitary circuits under unitary and open-system dynamics. **(a)** Power-law exponent $\alpha = 0.14$. The average operator size grows super-linearly under unitary dynamics (black), according to the power-law [17], $\overline{\mathcal{S}} \sim t^{1/(2\alpha-1)}$. Under open-system dynamics (red), the average operator size eventually reverts to sub-linear growth. **(b)** Power-law exponent $\alpha = 0.7$. The average operator size grows super-linearly under unitary dynamics (black), now according to a shrunk exponential [17], $\overline{\mathcal{S}} \sim e^{t^{\log 2 (1/\alpha)}}$. Under open-system dynamics (red), the average operator size eventually plateaus, similar to our observation in all-to-all interacting systems [Fig. 2(b) of the main text]. The arguments in the text predict that this apparent plateau is in fact a slow linear growth, however, it is difficult to precisely distinguish these the two scenarios here.

arrive at our final result:

$$I^{(2)}(A, E) = N + n_A + \log_2 \left[ \frac{1}{d d_A} \sum_{P_A} \left\langle \hat{P}_A(t) \hat{P}_A^\dagger(t) \right\rangle \right] \\ = 2 n_A + \log_2 \left[ \frac{1}{d_A^2} \sum_{P_A} \left\langle \hat{P}_A(t) \hat{P}_A^\dagger(t) \right\rangle \right]. \tag{37}$$

The argument of the logarithm is the average operator normalization over a complete basis of $d_A^2$ operators on subsystem $A$. We can easily check the limits on this equation. Under unitary dynamics, the average normalization is unity and so the mutual information is maximal, $I^{(2)}(A, E) = 2n_A$. Under a maximally decohering channel, the normalization approaches zero for all operators except the identity. This gives an average normalization of $1/d_A^2$, and thus a mutual information of zero, $I^{(2)}(A, E) = 2n_A - 2n_A = 0$.

## OPEN-SYSTEM OPERATOR GROWTH IN LONG-RANGE INTERACTING SYSTEMS

In this section, we turn to operator growth in long-range interacting open-system dynamics. We begin by summarizing previous results on operator growth under long-range interacting unitary dynamics. We then provide a brief intuitive picture to explain the effect of open-system dynamics, and explore these predictions numerically in random unitary circuits (Fig. 1).

Operator growth in long-range interacting dynamics has been studied in the context of random unitary circuits [16, 17] and biological physics [18, 19], and is known to display qualitatively distinct behavior compared to short-range dynamics. In particular, these works establish several regimes of operator growth, dependent on the power-law, $\alpha$, of the interaction and the spatial dimension, $d$. Here, long-range hops are designated to occur with a probability, $\sim 1/r^{2\alpha}$, we refer to the final section of the supplemental material for details. For $d + 1/2 < \alpha$, operator growth proceeds similarly to in short-range dynamics: the average operator size grows ballistically, $\mathcal{S} \sim t^d$, and the size width grows at a slower rate dominated by fluctuations at the light cone boundary. For $d < \alpha < d + 1/2$, the average operator size exhibits a *super-ballistic* power law in time, $\mathcal{S} \sim t^{\frac{1}{2\alpha-2d}}$. Moreover, fluctuations in the light cone boundary become the same order as the light cone width [18], implying that the size distribution becomes broad ($\delta \mathcal{S} \sim \overline{S}$). For $d/2 < \alpha < d$, the average operator size exhibits a stretched exponential, $\mathcal{S} \sim e^{t^{\log 2 (d/\alpha)}}$, again with large fluctuations at the boundary. Finally, for $\alpha < d/2$, the system is effectively all-to-all coupled and operators grow according a simple exponential in time.

In Ref. [18], all of the above patterns of growth are explained by a single remarkably simple model for operator growth. The central observation of this model is that the value of the light cone radius at time $t$, $r(t)$, is determined

primarily by contributions from long-range hops that were "seeded" from the bulk of the light cone at an earlier time, $t_{\text{seed}} \approx r^{-1}(r(t)/2)$. These seeds grow separately from the bulk of the light cone until the time of interest, $t$, at which the light cones of the seed and core combine. The radius $r(t)$ can then be solved for in a self-consistent approach, by demanding that the expected number of such hops up to the time $t$ [expressed in terms of $d$, $\alpha$, and $r(t)$] is equal to one. For more details, we refer to Ref. [18]. We note that this simple picture explains why the size distributions are broad (for power laws $\alpha < d + 1/2$). The long-range hops that contribute to the light cone boundary at time $t$ occur via a Poisson process with expected value 1. This leads to comparable fluctuations in the number of such hops; in particular, there will be some $\sim \mathcal{O}(1)$ probability that no hop of distance $r(t)$ has occurred at all. This is in sharp contrast to operator growth in short-range systems, where fluctuations in the light cone boundary receive equal contributions from fluctuations at all previous times [20].

A full analysis of long-range operator spreading in the presence of error would represent a substantial undertaking. We note that the manifestation of error that is relevant for time-evolution of quantum mechanical operators has no physical analog in the biological context where the previous studies of operator spreading originated [16–19]. Here, we instead propose a simple modification of the intuitive picture above. Consider two possible outcomes of the aforementioned Poisson process: one in which no hop of distance $r(t)$ occurs, and one in which $\geq 1$ such hops occur. Both outcomes occur with $\sim \mathcal{O}(1)$ probability under unitary dynamics. However, under open-system dynamics, the latter scenario will have be suppressed *exponentially* in the operator size of the seed integrated over time, i.e. proportional to $\sim \exp\left(-\varepsilon \int_{t_{\text{seed}}}^{t} dt' \, \mathcal{S}_{\text{seed}}(t')\right)$. We therefore expect that sufficiently long-range hops will cease to contribute to open-system operator growth, when this suppression becomes small. This implies that at sufficiently late times, operator growth will asymptote to ballistic behavior, $\mathcal{S} \sim t^d$, dominated by finite-range hops.

We explore these predictions by numerically simulating open-system long-range interacting RUCs for $\alpha = 0.7, 1.4$ (Fig. 1). For $\alpha = 1.4$, consistent with previous results, we observe that the average operator size grows super-ballistically under unitary dynamics (black). In contrast, under open-system dynamics (red), we indeed find that the average operator size eventually crosses over to sub-ballistic growth at late times. We observe qualitatively similar results for $\alpha = 0.7$. Namely, the average operator size grows consistently with a shrunk exponential at early times, and appears to plateau to a constant at late times. While we cannot firmly establish the functional form of the late time behavior in either case, the crossover from super-linear to sub-linear growth is consistent with the simple model outlined in the previous paragraph.

## OPEN-SYSTEM OPERATOR GROWTH IN FREE-FERMION INTEGRABLE SPIN CHAINS

In this section, we address operator growth in free-fermion integrable spin chains (FFISCs) with open-system dynamics (Fig. 2). Under unitary dynamics, FFISCs have been found to exhibit qualitatively different behavior of operator growth compared to chaotic spin chains [7, 21]. We will begin by showing that under unitary dynamics the size distributions of FFISCs are in fact *broad* ($\delta\mathcal{S} \sim \mathcal{S}$), unlike those of chaotic short-range 1D systems ($\delta\mathcal{S} \sim \sqrt{\mathcal{S}}$). We will then show that open-system dynamics eventually cause the average operator size of FFISCs to decrease in time, $\overline{\mathcal{S}} \sim 1/t$. In contrast to our other examples, we do not expect this behavior to be universal for all local error models. For instance, errors such as spontaneous emission will destroy the free fermion integrability of the model and are thus beyond the analytic framework we introduce below. Other errors, such as dephasing, will have a substantially weaker effect owing to the particular structure of time-evolved operators in FFISCs.

For concreteness, we focus on the XX model:

$$H = J \sum_{i=-\infty}^{\infty} (X_i X_{i+1} + Y_i Y_{i+1}) = J \sum_{i=-\infty}^{\infty} (r_i \ell_{i+1} + \ell_i r_{i+1}). \tag{38}$$

We expect the physics of other free-fermion integrable models to be similar. On the right side, we re-write the Pauli matrices, $X$ and $Y$, in terms of raising and lowering operators, $X_i = r_i + \ell_i$, $Y_i = i(r_i - \ell_i)$. We can solve this

Hamiltonian by defining fermion operators via the Jordan-Wigner transform:

$$r_i = \left(\prod_{0 \leq j < i}[2c_j^\dagger c_j - 1]\right) c_i^\dagger$$
$$\ell_i = \left(\prod_{0 \leq j < i}[2c_j^\dagger c_j - 1]\right) c_i \quad (39)$$
$$Z_i = 2c_i^\dagger c_i - 1.$$

The Hamiltonian becomes:

$$H = J \sum_{i=-\infty}^{\infty} \left(c_i^\dagger c_{i+1} + c_i c_{i+1}^\dagger\right), \quad (40)$$

corresponding to hopping on a one-dimensional chain of free fermions.

*Operator growth under unitary dynamics.*—The Jordan-Wigner transformation allows us to exactly solve for the time-evolution of operators that map to local fermion bilinears. We focus on one such example: the fermion number operator, $Z_0 = 2c_0^\dagger c_0 - 1$ (taken to lie at site $i = 0$). Under unitary dynamics, the Heisenberg equation is:

$$\partial_t Z_0(t) = i[Z_0(t), H]. \quad (41)$$

Note that repeated commutations $Z_0$ with the Hamiltonian give rise to operators exclusively of the form $r_x Z_{x+1} \ldots Z_{x'-1} \ell_{x'}$, for $x' \geq x$, and $\ell_{x'} Z_{x'+1} \ldots Z_{x-1} r_x$, for $x \geq x'$. In the fermion picture, these correspond to bilinears, $c_x^\dagger c_{x'}$. Making this explicit, we define the Pauli strings,

$$S_{x,x'} \equiv c_x^\dagger c_{x'} = \begin{cases} -P_x Z_{x+1} \ldots Z_{x'-1} M_{x'} & x < x' \\ +P_x M_x & x = x' \\ -M_{x'} Z_{x'+1} \ldots Z_{x-1} P_x & x > x' \end{cases}. \quad (42)$$

We can decompose the time-evolved operator as a sum of such strings,

$$Z_0(t) = 2 \sum_{x,x'} a_{x,x'}(t) \left(S_{x,x'} - \delta_{x,x} \mathbb{1}/2\right), \quad (43)$$

with coefficients $a_{x,x'}(t)$. Note that the identity factors guarantee that $\mathrm{tr}(Z_0(t)) = 0$. The Heisenberg equation becomes:

$$\sum_{x,x'} \partial_t a_{x,x'}(t) \left(S_{x,x'} - \delta_{x,x} \mathbb{1}/2\right) = i \sum_{x,x'} a_{x,x'}(t)[S_{x,x'}, H]$$
$$= iJ \sum_{x,x'} a_{x,x'}(t)[S_{x-1,x'} + S_{x+1,x'} - S_{x,x'-1} - S_{x,x'+1}] \quad (44)$$

which gives the following time-evolution for the coefficients, $a_{x,x'}(t)$:

$$\partial_t a_{x,x'}(t) = iJ[a_{x-1,x'}(t) + a_{x+1,x'}(t) - a_{x,x'-1}(t) - a_{x,x'+1}(t)]. \quad (45)$$

This is the equation of motion for two (decoupled) free fermions, indexed by $x$ and $x'$ respectively. It can be solved exactly via Fourier transform.

We now present the solution to this equation in the continuum limit, $a_{x,x'}(t) \to a(x, x'; t)$. This limit neglects lattice-scale fluctuations in the coefficients $a_{x,x'}(t)$, but we expect it to broadly capture the operators size distribution in the bulk of the light cone (see Ref. [21] for a full treatment of unitary operator growth without the continuum approximation). The Heisenberg equation becomes:

$$-i\partial_t a(x, x'; t) = J\partial_x^2 a(x, x'; t) - J\partial_{x'}^2 a(x, x'; t). \quad (46)$$

We can solve this via a Fourier transform with the initial condition, $a(x, x'; 0) = \delta(x)\delta(x')$:

$$a(x, x'; t) = \frac{1}{(2\pi)^2} \int dk \int dk' \, e^{-ikx} e^{-ik'x'} e^{-iJk^2 t} e^{iJk'^2 t}$$
$$= \frac{1}{(2\pi)2Jt} e^{-ix^2/4Jt} e^{ix'^2/4Jt}. \tag{47}$$

The solution has non-zero support at every site at any non-zero time, which is clearly unphysical. We can rectify this by smoothing the wavefunction over an initial Gaussian wavepacket of width $\sim 1$ (i.e. the lattice spacing):

$$\begin{aligned}
a_s(x, x'; t) &= \frac{1}{\sqrt{2\pi}} \int dy \int dy' \exp\left(-\frac{(x-y)^2}{2} - \frac{(x'-y')^2}{2}\right) a(y, y'; t) \\
&= \frac{1}{\sqrt{2\pi}} \frac{1}{(2\pi)2Jt} \int dy \int dy' \exp\left(-\frac{(x-y)^2}{2} - i\frac{y^2}{4Jt} - \frac{(x'-y')^2}{2} + i\frac{y'^2}{4Jt}\right) \\
&= \frac{1}{\sqrt{2\pi}} \frac{1}{(2\pi)2Jt} \int dy \int dy' \exp\left(-\frac{y^2(1 + i/2Jt)}{2} - yx - \frac{x^2}{2} - \frac{y'^2(1 - i/2Jt)}{2} - y'x' - \frac{x'^2}{2}\right) \\
&= \frac{1}{\sqrt{2\pi} 2Jt\sqrt{(1 + i/2Jt)(1 - i/2Jt)}} \exp\left(-\frac{x^2}{2}\left[1 - \frac{1}{1 + i/2Jt}\right] - \frac{x'^2}{2}\left[1 - \frac{1}{1 - i/2Jt}\right]\right) \\
&= \frac{1}{\sqrt{2\pi}\sqrt{(2Jt)^2 + 1}} \exp\left(-\frac{x^2}{2(1 - i2Jt)}\right) \exp\left(-\frac{x'^2}{2(1 + i2Jt)}\right).
\end{aligned} \tag{48}$$

This gives a normalized probability distribution for $x, x'$,

$$\begin{aligned}
|a_s(x, x'; t)|^2 &= \frac{1}{2\pi[(2Jt)^2 + 1]} \exp\left(-\frac{x^2}{2(1 - i2Jt)} - \frac{x^2}{2(1 + i2Jt)}\right) \exp\left(-\frac{x'^2}{2(1 + i2Jt)} - \frac{x'^2}{2(1 - i2Jt)}\right) \\
&= \frac{1}{2\pi[(2Jt)^2 + 1]} \exp\left(-\frac{x^2/2}{1 + (2Jt)^2}\right) \exp\left(-\frac{x'^2/2}{1 + (2Jt)^2}\right) \\
&\approx \frac{1}{2\pi(2Jt)^2} \exp\left(-\frac{1}{2}(x/2Jt)^2\right) \exp\left(-\frac{1}{2}(x'/2Jt)^2\right),
\end{aligned} \tag{49}$$

corresponding to a Gaussian centered about $x = x' = 0$, with a ballistically growing width, $2Jt$.

To compute the operator size distribution, note that the size superoperator acts on each string, $S_{x,x'}$, as:

$$\mathcal{S}\{S_{x,x'}\} = (|x - x'| + 1)S_{x,x'} - \frac{1}{2}\mathbb{1}\delta_{x,x'}. \tag{50}$$

This implies that $S_{x,x'} - \frac{1}{2}\mathbb{1}\delta_{x,x'}$ is a size eigenstate with size $(|x-x'|+1)$. The size distribution of $Z_0(t)$ is therefore:

$$P(\mathcal{S}) = \sum_{\{x,x':|x-x'|+1=\mathcal{S}\}} |a_{x,x'}(t)|^2 \approx \int dx \left(|a_s(x, x + \mathcal{S} - 1; t)|^2 + |a_s(x, x - \mathcal{S} + 1; t)|^2\right) \tag{51}$$

Changing coordinates, $\delta H = x + x'$, $\delta = x - x'$, the operator wavefunction becomes:

$$|a_s(\delta H, \delta; t)|^2 \approx \frac{1}{2(2\pi)(2Jt)^2} \exp\left(-\frac{1}{4}(\delta H/2Jt)^2\right) \exp\left(-\frac{1}{4}(\delta/2Jt)^2\right), \tag{52}$$

which gives a size distribution:

$$\begin{aligned}
P(\mathcal{S}) &\approx \int d\delta H \, |a_s(\delta H, \mathcal{S} - 1; t)|^2 + |a_s(\delta H, -\mathcal{S} + 1; t)|^2 \\
&= \frac{1}{2\sqrt{\pi}Jt} \exp\left(-\frac{1}{4}\left(\frac{\mathcal{S} - 1}{2Jt}\right)^2\right). \\
&\approx \frac{1}{2\sqrt{\pi}Jt} \exp\left(-\frac{1}{4}\left(\frac{\mathcal{S}}{2Jt}\right)^2\right).
\end{aligned} \tag{53}$$

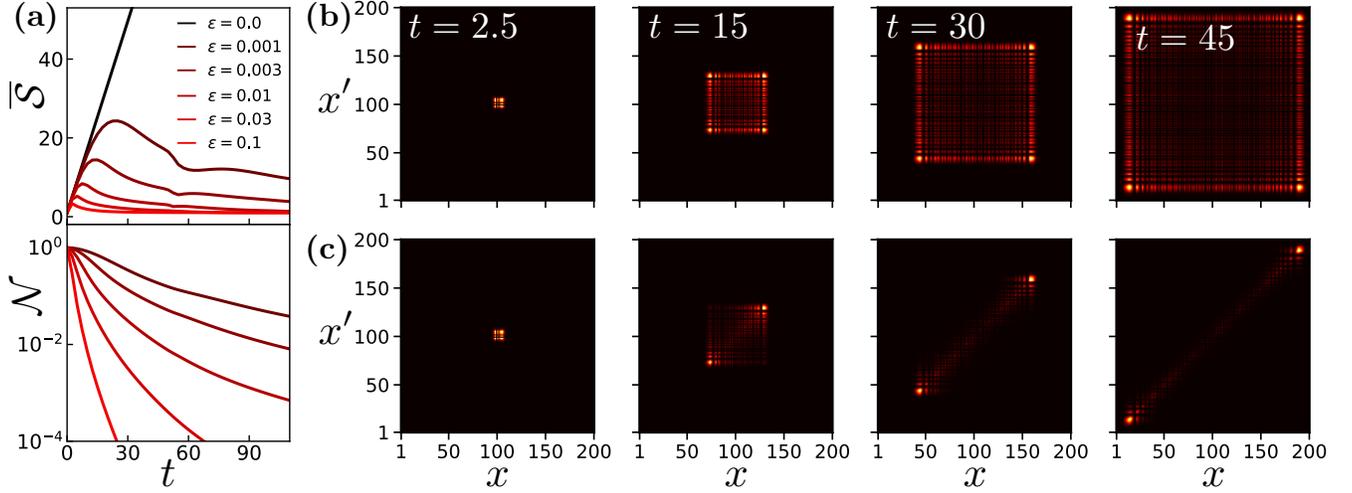

FIG. 2. Operator growth in free-fermion integrable spin chains with isotropic single-qubit decoherence [Eq. (58)]. **(a)** The average operator size increases linearly at early times, and before inverting to decay towards one at a time-scale set by the error rate, $\varepsilon$. The Loschmidt echo fidelity (i.e. the operator normalization) displays non-concave behavior as a result of this reverse in operator growth. **(b)** The probability distribution, $a_{x,x'}(t)$, for the location of the endpoints, $x, x'$, of a time-evolved operators' Pauli strings under unitary dynamics. Each endpoint spreads ballistically in either direction, and the interior of the distribution is approximately uniform. **(c)** The same probability distribution under open-system dynamics with $\varepsilon = 0.003$. The small-size components (which lie along the diagonal, $x \approx x'$) are nearly unaffected by error, while large-size components (which lie highly off-diagonal) are strongly damped. Note that the time of the final image, $t = 45$, occurs shortly before the first kink appearing in **(a)**.

The size distribution is the positive half of a Gaussian with mean zero and width $2\sqrt{2}Jt$. The average size thus grows ballistically,

$$\mathcal{S} \approx \frac{4Jt}{\sqrt{\pi}}, \tag{54}$$

and the size width likewise,

$$\delta \mathcal{S} \approx 2\sqrt{2}\sqrt{1 - 2/\pi}Jt. \tag{55}$$

This is in sharp contrast with chaotic spin chains, which exhibit diffusive growth of the size width, $\delta\mathcal{S} \sim \sqrt{\mathcal{S}}$.

*Operator growth under open-system dynamics.—*We now turn to the effect of open-system dynamics on operator growth. For simplicity, we study isotropic decoherence on each qubit, in which case the latter portion of the Lindbladian is directly given by the operator size. Again, in contrast to chaotic systems, we expect that different forms of open-system dynamics will affect FFISCs in different ways, since the large-size Pauli strings composing $\hat{Z}_0(t)$ contain nearly entirely Pauli-$Z$ components [Eq. (42)]. For example, this implies that dephasing dynamics will have a nearly negligible effect on time-evolution. On the other hand, open-system dynamics such as spontaneous emission will evolve $\hat{Z}_0(t)$ outside the given set of strings, breaking free fermion integrability.

The Heisenberg equation becomes:

$$\partial_t Z_0(t) = i[Z_0(t), H] - \varepsilon \mathcal{S}\{Z_0(t)\}. \tag{56}$$

Using the expansion in Eq. (43), we have:

$$\sum_{x,x'} \partial_t a_{x,x'}(t) \left(S_{x,x'} - \delta_{x,x} \mathbb{1}/2\right) = \sum_{x,x'} a_{x,x'}(t) \left(i[S_{x,x'}, H] - \varepsilon \mathcal{S}\{S_{x,x'}\}\right)$$
$$= \sum_{x,x'} a_{x,x'}(t) \left(iJ[S_{x-1,x'} + S_{x+1,x'} - S_{x,x'-1} - S_{x,x'+1}] - \varepsilon(|x - x'| + 1)\left(S_{x,x'} - \delta_{x,x} \mathbb{1}/2\right)\right), \tag{57}$$

which gives the following time-evolution for the Pauli string coefficients:

$$\partial_t a_{x,x'}(t) = iJ[a_{x-1,x'}(t) + a_{x+1,x'}(t) - a_{x,x'-1}(t) - a_{x,x'+1}(t)] - \varepsilon(|x - x'| + 1)a_{x,x'}(t). \tag{58}$$

Note that the coordinates, $x$ and $x'$, are now coupled by the Lindbladian, since the error is proportional the absolute value of the separation between the two coordinates (i.e. the size).

To approximately solve this equation, first recall that in the fermion picture for unitary time-evolution, the growth of $Z_0(t)$ corresponds a fermionic creation and annihilation operator each dispersing independently. At large times, $t \gg 1/J$, each momentum-component of each fermion will propagate independently of all other momenta (formally, this corresponds to a saddle point approximation when solving Eq. (44) via a Fourier transform, see Ref. [21]). For a given momentum $k$, the group velocity, $v(k) = \partial_k \epsilon(k)$, is determined by the dispersion relation, $\epsilon(k)$, of the Hamiltonian (for the XX model, we have $\epsilon(k) = 2J\cos(k) \approx 2J + Jk^2$ in the continuum limit). Turning to Lindbladian time-evolution, we now consider a pair of momenta, $k, k'$, for the creation and annihilation operator respectively. The size of the Pauli strings corresponding to these momenta grows linearly in time, according to:

$$\mathcal{S}_{k,k'}(t) = |v(k) - v(k')|t. \tag{59}$$

We therefore expect normalization of the particular pair of momenta to decay by a factor,

$$\sim \exp\left(-2\varepsilon \int_0^t dt' \mathcal{S}_{k,k'}(t)\right) = \exp\left(-\varepsilon t \mathcal{S}_{k,k'}(t)\right). \tag{60}$$

Integrating over all momenta pairs gives a size distribution,

$$P(\mathcal{S}) \approx \frac{1}{2\sqrt{\pi}Jt} \exp\left(-\frac{1}{4}\left(\frac{\mathcal{S}}{2Jt}\right)^2 - \varepsilon \mathcal{S} t\right). \tag{61}$$

We expect this treatment to hold when $\varepsilon/J \ll 1$, which guarantees that the effect of error, $\sim \varepsilon J t^2$, will be negligible until sufficiently large times ($t \gg 1/J$) that our approximation holds.

The normalization of the operator size distribution is:

$$\mathcal{N} = \int_0^\infty d\mathcal{S}\, P(\mathcal{S}) = \exp(4J^2\varepsilon^2 t^4) \cdot \mathrm{erfc}\left(2J\varepsilon t^2\right), \tag{62}$$

where the complementary error function is defined as, $\mathrm{erfc}(x) \equiv (2/\sqrt{\pi}) \int_x^\infty e^{-x^2}$. The average size can also be computed exactly in terms of the error function, although we do not express it here. At late times ($J\varepsilon t^2 \gg 1$), the size distribution asymptotes to,

$$P(\mathcal{S}) \to \frac{1}{2\sqrt{\pi}Jt} \exp\left(-\varepsilon \mathcal{S} t\right), \tag{63}$$

which gives an algebraic decay for the normalization,

$$\mathcal{N} \to \frac{1}{2\sqrt{\pi}J\varepsilon t^2}. \tag{64}$$

This can be derived from Eq. (62) using the approximation, $\mathrm{erfc}(x) \approx e^{-x^2}/(\sqrt{\pi}x)$, for the complementary error function at large $x$. This normalization decay implies a $\sim 1/t$ decay of the average operator size,

$$\mathcal{S} = \frac{-\partial_t \log(\mathcal{N})}{2\varepsilon} \to \frac{1}{\varepsilon t}, \tag{65}$$

as quoted in Table 1 of the main text.

To explore open-system operator growth in FFISCs outside the continuum limit, we numerically solve Eq. (58) for $N = 200$ with open boundary conditions. As shown in Fig. 2, we observe that the average operator size indeed increases linearly at early times and decays towards its minimal value, one, at late times. We also observe "kinks" in the time profile of the average operator size. By analyzing the full Pauli string distribution (Fig. 2), $a_{x,x'}(t)$, we see that these kinks occur precisely when fermion operators reflect off the boundary of the finite-size system.

# NUMERICAL SIMULATION DETAILS

In this section we provide details on the numerical simulations presented in Figs. 2, 3 of the main text and Figs. 1, 2 of the supplemental material.

*Haar-random unitary circuits.*—We begin with the random unitary circuit (RUC) simulations presented in Fig. 2 of the main text and Fig. 1 of the supplemental material. Haar-random unitary circuits are toy models for operator spreading, in which averages of quantities that contain two or fewer copies of a time-evolved operator are efficiently simulable by averaging instead over Clifford-random unitaries [20, 22–24]. Note that the generating function for the size distribution (Fig. 4 of the main text) contains two copies of $\hat{M}(t)$ can thus be efficiently simulated. The zeroth and first moments of the size distribution provide the Loschmidt echo fidelity and the average operator size. The RUC formalism can be also adapted to incorporate *open-system* operator spreading [4]. Below, we do so for the specific error model in Eq. (4) of the main text. As discussed earlier in the supplemental material, this model exactly captures the effect of single-qubit Pauli errors, and also applies to leading order in the error rate for arbitrary single-qubit errors (assuming in both cases that errors occur at the same rate at each qubit, although the model is easily modified if this is not the case).

We consider three RUC models: (i) 1D short-range interacting [20], (ii) 1D long-range interacting [16, 17], and (iii) 0D all-to-all interacting [14]. In each model, a single time step, $\delta t$, consists of choosing a pair of qubits and acting on this pair with a Haar-random unitary, and applying the global quantum channel, $\exp(-\varepsilon \mathcal{S} \delta t)$. A unit of physical time consists of $N$ iterations of this procedure, so that each qubit is acted on by two unitaries per unit time, on average. The pairs are chosen as follows: (i) in the short-range 1D RUC, a qubit $j$ is sampled randomly and paired with its nearest neighbor, $j+1$, (ii) in the long-range 1D RUC, a pair of qubits $i$ and $j$ are sampled according to a power law distribution in their distance, $\sim 1/|i-j|^{2\alpha}$ [16, 17], and (iii) in the 0D RUC, pairs are sampled entirely randomly. In cases (i) and (ii) we take periodic boundary conditions.

We simulate the average generating function of the operator size distribution by tracking the Clifford time-evolution of a local Pauli operator of interest, and sampling over realizations of Clifford-random unitaries [20, 24]. In each Clifford realization, $\mathcal{C}$, the time-evolved operator, $\hat{M}_\mathcal{C}(t)$, has a well-defined size at every time step (since Pauli operators are eigenstates of the size superoperator). Under open-system dynamics, the individual Pauli operator $\hat{M}_\mathcal{C}(t)$ thus evolves exactly as:

$$\hat{M}_\mathcal{C}(t) = \exp\left(-\varepsilon \int_0^t dt' \mathcal{S}_\mathcal{C}(t')\right) \cdot \hat{M}_{\mathcal{C},U}(t). \tag{66}$$

Here, $\hat{M}_{\mathcal{C},U}(t)$ represents the operator time-evolved by Clifford unitaries under *unitary* dynamics (i.e. $\varepsilon = 0$), and $\mathcal{S}_\mathcal{C}(t')$ is the size of this operator at time $t'$. The average generating function of the size distribution over Haar-random realizations, $\mathcal{U}$, is therefore equal to:

$$G_\mathcal{S}(\mu) = \frac{1}{N_\mathcal{C}} \sum_\mathcal{C} \left\langle \hat{M}_\mathcal{C}(t) e^{-\mu \mathcal{S}} \left\{ \hat{M}_\mathcal{C}(t) \right\} \right\rangle = \frac{1}{N_\mathcal{C}} \sum_\mathcal{C} \mathcal{N}_\mathcal{C}(t) e^{-\mu \mathcal{S}_\mathcal{C}(t)}, \tag{67}$$

where we denote the normalization of a single Clifford realization as:

$$\mathcal{N}_\mathcal{C}(t) = \exp\left(-2\varepsilon \int_0^t dt' \mathcal{S}_\mathcal{C}(t')\right). \tag{68}$$

The average generating function is a sum of the generating functions over each Clifford realization (i.e. the simple exponentials $e^{-\mu \mathcal{S}_\mathcal{C}(t)}$), weighted by its normalization, $\mathcal{N}_\mathcal{C}(t)$. From Eq. (67), we obtain the average Loschmidt echo fidelity,

$$\mathcal{N}_t = \frac{1}{N_\mathcal{C}} \sum_\mathcal{C} \mathcal{N}_\mathcal{C}(t), \tag{69}$$

and the moments of the size distribution,

$$\overline{\mathcal{S}^n} = \frac{\frac{1}{N_\mathcal{C}} \sum_\mathcal{C} \mathcal{N}_\mathcal{C}(t) \left(\mathcal{S}^\mathcal{C}(t)\right)^n}{\mathcal{N}_t}. \tag{70}$$

In Fig. 2 of the main text, we plot our phenomenological predictions as dashed lines in addition to data from numerical simulations. For Fig. 2(a) [i.e. model (i)], the dashed curve is given by $\overline{S} = 2v_B t - \varepsilon c t^2/2$, with the butterfly velocity, $v_B = (3/5)(3/4) = 9/20$, set by theory [20], and the size width parameter, $c = 1.14$, fit to the average operator size data via linear regression. For Fig. 2(b) [i.e. model (iii)], the dashed curve is given by $\overline{S} = \frac{e^{\lambda t}}{1 + e^{\lambda t}/S_p}$, with the value of the plateau size, $S_p = \frac{\lambda}{b^2 \varepsilon + 4\lambda/3N}$, and the Lyapunov exponent, $\lambda = 2(3/5) = 6/5$, are set by theory, and the size width parameter, $b = 1.00$, is fit to the average operator size via linear regression. This is the solution to the phenomenological size growth equation, $\partial_t \overline{S} = \lambda \overline{S} - (\varepsilon b^2 + \frac{4\lambda}{3N})\overline{S}^2$, obtained from the equation in the main text by setting $\delta S = bS$ and including a $1/N$ correction to ensure that under unitary dynamics the average operator size saturates to the late time value, $3N/4$.

*Hamiltonian dynamics.*—In Fig. 3 of the main text, we compute the OTOCs of local operators in the scenario in which forwards and backwards time-evolution are both performed unitarily, but under unequal Hamiltonians. Specifically, we consider the following spin chain Hamiltonians:

$$\hat{H}_1 = \hat{H} - \frac{\eta}{2}\delta\hat{H}, \quad \hat{H}_2 = \hat{H} + \frac{\eta}{2}\delta\hat{H}. \tag{71}$$

Here, the Hamiltonian $\hat{H}$ is given by:

$$\hat{H} = \sum_{i=1}^{N} h_z \hat{Z}_i + \sum_{i=1}^{N-1} \left(J_x \hat{X}_i \hat{X}_{i+1} + J_y \hat{Y}_i \hat{Y}_{i+1} + J_z \hat{Z}_i \hat{Z}_{i+1}\right) + \sum_{i=1}^{N-2} \left(J'_x \hat{X}_i \hat{X}_{i+2} + J'_y \hat{Y}_i \hat{Y}_{i+2} + J'_z \hat{Z}_i \hat{Z}_{i+2}\right), \tag{72}$$

which consists of single-site fields, $h_z = 0.4$, strong nearest-neighbor interactions, $J_x = 0.6, J_y = -0.8, J_z = 1.1$, and weak next-nearest-neighbor interactions, $J'_x = -0.3, J'_y = 0.3, J'_z = 0.15$. Here, we choose "generic" values for the interaction strengths in order to avoid behavior associated with closeness to integrable points (for example, the Heisenberg point at $J_\alpha = J, J'_\alpha = 0$). We do not otherwise expect our results to depend on the specific values chosen. The perturbing Hamiltonian,

$$\eta \, \delta \hat{H} = \sum_i \left( h_{x,i} \hat{X}_i + h_{y,i} \hat{Y}_i \right), \tag{73}$$

consists of disordered single-site fields in the perpendicular directions, with each magnitude drawn randomly from the interval $[-\eta, \eta]$ with $\eta = 0.4$. This choice of Hamiltonians possesses one special feature, namely that the perturbing Hamiltonian has zero overlap with all powers of the average Hamiltonian, $\text{tr}\left(\delta \hat{H} \, \hat{H}^m\right) = 0$ for all $m$. As discussed earlier in the supplemental material, we expect errors with non-zero overlap to contribute subleading corrections (suppressed by powers of $\sim 1/t$) to our picture of operator spreading.

We numerically compute the Loschmidt echo fidelity,

$$\mathcal{N}(t) = \langle e^{-iH_1 t} \hat{M} e^{iH_1 t} e^{-iH_2 t} \hat{M} e^{iH_2 t} \rangle, \tag{74}$$

and the OTOCs,

$$C(t; \hat{P}_i) = \langle e^{-iH_1 t} \hat{M} e^{iH_1 t} \hat{P}_i e^{-iH_2 t} \hat{M} e^{iH_2 t} \hat{P}_i \rangle \tag{75}$$

of the above Hamiltonians via Krylov-subspace methods for $N = 12$ and $10^4$ disorder realizations. Note that Loschmidt echo fidelity, $\mathcal{N}(t)$, is not guaranteed to be positive in this scenario (although we expect it to be in the many-body limit). Indeed, at late times where $\mathcal{N}(t)$ is expected to be exponentially small, we observe that the value of $\mathcal{N}(t)$ for individual disorder realizations may take small negative values. We find that the average of $\mathcal{N}(t)$ over disorder realizations is significantly more stable and remains positive for all times considered.

In Fig. 3 of the main text, we plot the average of the OTOC over both disorder realizations and the set of single-qubit Pauli operators, $\hat{P}_i \in \{\mathbb{1}_i, X_i, Y_i, Z_i\}$, divided by the average Loschmidt echo fidelity over disorder realizations. We comment that, if the experiments we wish to model contain only a single realization of the perturbing Hamiltonian, it would be more appropriate to divide the OTOC by the Loschmidt echo fidelity *before* averaging over disorder realizations. However, we find that this leads to numerical instabilities at the small system sizes we consider. In Fig. 3(a) we take the initial operator $\hat{M} = \hat{X}_0$, which does not overlap the Hamiltonian, while in Fig. 3(b) we take the initial operator $\hat{M} = \hat{Z}_0$, which has non-zero overlap.

*Free fermions integrable spin chains.*—Time-evolution for Fig. 2 is performed by exactly solving Eq. (58) via matrix exponentiation, with open boundary condition for $J = 1$ and $N = 200$.



\end{document}